\documentclass[a4paper, traditabstract,longauth]{aa} 

\usepackage{graphicx,amsmath}
\usepackage{txfonts}
\usepackage{lscape}
\usepackage{color}
\usepackage{amssymb}
\usepackage{booktabs}
\usepackage{subfig}
\usepackage[breaklinks, colorlinks, citecolor=cyan]{hyperref}
\usepackage{natbib}

\def\setsymbol#1#2{\expandafter\def\csname #1\endcsname{#2}}
\def\getsymbol#1{\csname #1\endcsname}

\def\Planck{{\it Planck\/}}

\newbox\tablebox    \newdimen\tablewidth
\def\leaderfil{\leaders\hbox to 5pt{\hss.\hss}\hfil}
\def\endPlancktable{\tablewidth=\columnwidth 
    $$\hss\copy\tablebox\hss$$
    \vskip-\lastskip\vskip -2pt}
\def\endPlancktablewide{\tablewidth=\textwidth 
    $$\hss\copy\tablebox\hss$$
    \vskip-\lastskip\vskip -2pt}
\def\tablenote#1 #2\par{\begingroup \parindent=0.8em
    \abovedisplayshortskip=0pt\belowdisplayshortskip=0pt
    \noindent
    $$\hss\vbox{\hsize\tablewidth \hangindent=\parindent \hangafter=1 \noindent
    \hbox to \parindent{\sup{\rm #1}\hss}\strut#2\strut\par}\hss$$
    \endgroup}
\def\doubleline{\vskip 3pt\hrule \vskip 1.5pt \hrule \vskip 5pt}

\def\L2{\ifmmode L_2\else $L_2$\fi}

\def\DeltaT{\ifmmode \Delta T\else $\Delta T$\fi}
\def\deltat{\ifmmode \Delta t\else $\Delta t$\fi}
\def\fknee{\ifmmode f_{\rm knee}\else $f_{\rm knee}$\fi}
\def\Fmax{\ifmmode F_{\rm max}\else $F_{\rm max}$\fi}
\def\solar{\ifmmode{\rm M}_{\mathord\odot}\else${\rm M}_{\mathord\odot}$\fi}

\def\inv{\ifmmode^{-1}\else$^{-1}$\fi}
\def\mo{\ifmmode^{-1}\else$^{-1}$\fi}
\def\sup#1{\ifmmode ^{\rm #1}\else $^{\rm #1}$\fi}
\def\expo#1{\ifmmode \times 10^{#1}\else $\times 10^{#1}$\fi}
\def\,{\thinspace}
\def\lsim{\mathrel{\raise .4ex\hbox{\rlap{$<$}\lower 1.2ex\hbox{$\sim$}}}}
\def\gsim{\mathrel{\raise .4ex\hbox{\rlap{$>$}\lower 1.2ex\hbox{$\sim$}}}}

\def\simprop{\mathrel{\raise .4ex\hbox{\rlap{$\propto$}\lower 1.2ex\hbox{$\sim$}}}}
\def\deg{\ifmmode^\circ\else$^\circ$\fi}
\def\pdeg{\ifmmode $\setbox0=\hbox{$^{\circ}$}\rlap{\hskip.11\wd0 .}$^{\circ}
          \else \setbox0=\hbox{$^{\circ}$}\rlap{\hskip.11\wd0 .}$^{\circ}$\fi}
\def\arcs{\ifmmode {^{\scriptstyle\prime\prime}}
          \else $^{\scriptstyle\prime\prime}$\fi}
\def\arcm{\ifmmode {^{\scriptstyle\prime}}
          \else $^{\scriptstyle\prime}$\fi}
\newdimen\sa  \newdimen\sb
\def\parcs{\sa=.07em \sb=.03em
     \ifmmode \hbox{\rlap{.}}^{\scriptstyle\prime\kern -\sb\prime}\hbox{\kern -\sa}
     \else \rlap{.}$^{\scriptstyle\prime\kern -\sb\prime}$\kern -\sa\fi}
\def\parcm{\sa=.08em \sb=.03em
     \ifmmode \hbox{\rlap{.}\kern\sa}^{\scriptstyle\prime}\hbox{\kern-\sb}
     \else \rlap{.}\kern\sa$^{\scriptstyle\prime}$\kern-\sb\fi}
\def\ra[#1 #2 #3.#4]{#1\sup{h}#2\sup{m}#3\sup{s}\llap.#4}
\def\dec[#1 #2 #3.#4]{#1\deg#2\arcm#3\arcs\llap.#4}
\def\deco[#1 #2 #3]{#1\deg#2\arcm#3\arcs}
\def\rra[#1 #2]{#1\sup{h}#2\sup{m}}

\def\dots{\relax\ifmmode \ldots\else $\ldots$\fi}
\def\WHzsr{\ifmmode $W\,Hz\mo\,sr\mo$\else W\,Hz\mo\,sr\mo\fi}
\def\mHz{\ifmmode $\,mHz$\else \,mHz\fi}
\def\GHz{\ifmmode $\,GHz$\else \,GHz\fi}
\def\mKs{\ifmmode $\,mK\,s$^{1/2}\else \,mK\,s$^{1/2}$\fi}
\def\muKs{\ifmmode \,\mu$K\,s$^{1/2}\else \,$\mu$K\,s$^{1/2}$\fi}
\def\muKRJs{\ifmmode \,\mu$K$_{\rm RJ}$\,s$^{1/2}\else \,$\mu$K$_{\rm RJ}$\,s$^{1/2}$\fi}
\def\muKHz{\ifmmode \,\mu$K\,Hz$^{-1/2}\else \,$\mu$K\,Hz$^{-1/2}$\fi}
\def\MJysr{\ifmmode \,$MJy\,sr\mo$\else \,MJy\,sr\mo\fi}
\def\MJysrmK{\ifmmode \,$MJy\,sr\mo$\,mK$_{\rm CMB}\mo\else \,MJy\,sr\mo\,mK$_{\rm CMB}\mo$\fi}
\def\microns{\ifmmode \,\mu$m$\else \,$\mu$m\fi}

\def\muK{\ifmmode \,\mu$K$\else \,$\mu$\hbox{K}\fi}
\def\microK{\ifmmode \,\mu$K$\else \,$\mu$\hbox{K}\fi}
\def\muW{\ifmmode \,\mu$W$\else \,$\mu$\hbox{W}\fi}
\def\kms{\ifmmode $\,km\,s$^{-1}\else \,km\,s$^{-1}$\fi}
\def\kmsMpc{\ifmmode $\,\kms\,Mpc\mo$\else \,\kms\,Mpc\mo\fi}
\setsymbol{LFI:center:frequency:70GHz:units}{70.3\,GHz}
\setsymbol{LFI:center:frequency:44GHz:units}{44.1\,GHz}
\setsymbol{LFI:center:frequency:30GHz:units}{28.5\,GHz}

\setsymbol{LFI:center:frequency:70GHz}{70.3}
\setsymbol{LFI:center:frequency:44GHz}{44.1}
\setsymbol{LFI:center:frequency:30GHz}{28.5}

\setsymbol{LFI:center:frequency:LFI18:Rad:M:units}{71.7\GHz}
\setsymbol{LFI:center:frequency:LFI19:Rad:M:units}{67.5\GHz}
\setsymbol{LFI:center:frequency:LFI20:Rad:M:units}{69.2\GHz}
\setsymbol{LFI:center:frequency:LFI21:Rad:M:units}{70.4\GHz}
\setsymbol{LFI:center:frequency:LFI22:Rad:M:units}{71.5\GHz}
\setsymbol{LFI:center:frequency:LFI23:Rad:M:units}{70.8\GHz}
\setsymbol{LFI:center:frequency:LFI24:Rad:M:units}{44.4\GHz}
\setsymbol{LFI:center:frequency:LFI25:Rad:M:units}{44.0\GHz}
\setsymbol{LFI:center:frequency:LFI26:Rad:M:units}{43.9\GHz}
\setsymbol{LFI:center:frequency:LFI27:Rad:M:units}{28.3\GHz}
\setsymbol{LFI:center:frequency:LFI28:Rad:M:units}{28.8\GHz}
\setsymbol{LFI:center:frequency:LFI18:Rad:S:units}{70.1\GHz}
\setsymbol{LFI:center:frequency:LFI19:Rad:S:units}{69.6\GHz}
\setsymbol{LFI:center:frequency:LFI20:Rad:S:units}{69.5\GHz}
\setsymbol{LFI:center:frequency:LFI21:Rad:S:units}{69.5\GHz}
\setsymbol{LFI:center:frequency:LFI22:Rad:S:units}{72.8\GHz}
\setsymbol{LFI:center:frequency:LFI23:Rad:S:units}{71.3\GHz}
\setsymbol{LFI:center:frequency:LFI24:Rad:S:units}{44.1\GHz}
\setsymbol{LFI:center:frequency:LFI25:Rad:S:units}{44.1\GHz}
\setsymbol{LFI:center:frequency:LFI26:Rad:S:units}{44.1\GHz}
\setsymbol{LFI:center:frequency:LFI27:Rad:S:units}{28.5\GHz}
\setsymbol{LFI:center:frequency:LFI28:Rad:S:units}{28.2\GHz}

\setsymbol{LFI:center:frequency:LFI18:Rad:M}{71.7}
\setsymbol{LFI:center:frequency:LFI19:Rad:M}{67.5}
\setsymbol{LFI:center:frequency:LFI20:Rad:M}{69.2}
\setsymbol{LFI:center:frequency:LFI21:Rad:M}{70.4}
\setsymbol{LFI:center:frequency:LFI22:Rad:M}{71.5}
\setsymbol{LFI:center:frequency:LFI23:Rad:M}{70.8}
\setsymbol{LFI:center:frequency:LFI24:Rad:M}{44.4}
\setsymbol{LFI:center:frequency:LFI25:Rad:M}{44.0}
\setsymbol{LFI:center:frequency:LFI26:Rad:M}{43.9}
\setsymbol{LFI:center:frequency:LFI27:Rad:M}{28.3}
\setsymbol{LFI:center:frequency:LFI28:Rad:M}{28.8}
\setsymbol{LFI:center:frequency:LFI18:Rad:S}{70.1}
\setsymbol{LFI:center:frequency:LFI19:Rad:S}{69.6}
\setsymbol{LFI:center:frequency:LFI20:Rad:S}{69.5}
\setsymbol{LFI:center:frequency:LFI21:Rad:S}{69.5}
\setsymbol{LFI:center:frequency:LFI22:Rad:S}{72.8}
\setsymbol{LFI:center:frequency:LFI23:Rad:S}{71.3}
\setsymbol{LFI:center:frequency:LFI24:Rad:S}{44.1}
\setsymbol{LFI:center:frequency:LFI25:Rad:S}{44.1}
\setsymbol{LFI:center:frequency:LFI26:Rad:S}{44.1}
\setsymbol{LFI:center:frequency:LFI27:Rad:S}{28.5}
\setsymbol{LFI:center:frequency:LFI28:Rad:S}{28.2}

\setsymbol{LFI:white:noise:sensitivity:70GHz:units}{134.7\muKs}
\setsymbol{LFI:white:noise:sensitivity:44GHz:units}{164.7\muKs}
\setsymbol{LFI:white:noise:sensitivity:30GHz:units}{143.4\muKs}

\setsymbol{LFI:white:noise:sensitivity:70GHz}{134.7}
\setsymbol{LFI:white:noise:sensitivity:44GHz}{164.7}
\setsymbol{LFI:white:noise:sensitivity:30GHz}{143.4}

\setsymbol{LFI:white:noise:sensitivity:LFI18:Rad:M:units}{512.0\muKs}
\setsymbol{LFI:white:noise:sensitivity:LFI19:Rad:M:units}{581.4\muKs}
\setsymbol{LFI:white:noise:sensitivity:LFI20:Rad:M:units}{590.8\muKs}
\setsymbol{LFI:white:noise:sensitivity:LFI21:Rad:M:units}{455.2\muKs}
\setsymbol{LFI:white:noise:sensitivity:LFI22:Rad:M:units}{492.0\muKs}
\setsymbol{LFI:white:noise:sensitivity:LFI23:Rad:M:units}{507.7\muKs}
\setsymbol{LFI:white:noise:sensitivity:LFI24:Rad:M:units}{462.2\muKs}
\setsymbol{LFI:white:noise:sensitivity:LFI25:Rad:M:units}{413.6\muKs}
\setsymbol{LFI:white:noise:sensitivity:LFI26:Rad:M:units}{478.6\muKs}
\setsymbol{LFI:white:noise:sensitivity:LFI27:Rad:M:units}{277.7\muKs}
\setsymbol{LFI:white:noise:sensitivity:LFI28:Rad:M:units}{312.3\muKs}
\setsymbol{LFI:white:noise:sensitivity:LFI18:Rad:S:units}{465.7\muKs}
\setsymbol{LFI:white:noise:sensitivity:LFI19:Rad:S:units}{555.6\muKs}
\setsymbol{LFI:white:noise:sensitivity:LFI20:Rad:S:units}{623.2\muKs}
\setsymbol{LFI:white:noise:sensitivity:LFI21:Rad:S:units}{564.1\muKs}
\setsymbol{LFI:white:noise:sensitivity:LFI22:Rad:S:units}{534.4\muKs}
\setsymbol{LFI:white:noise:sensitivity:LFI23:Rad:S:units}{542.4\muKs}
\setsymbol{LFI:white:noise:sensitivity:LFI24:Rad:S:units}{399.2\muKs}
\setsymbol{LFI:white:noise:sensitivity:LFI25:Rad:S:units}{392.6\muKs}
\setsymbol{LFI:white:noise:sensitivity:LFI26:Rad:S:units}{418.6\muKs}
\setsymbol{LFI:white:noise:sensitivity:LFI27:Rad:S:units}{302.9\muKs}
\setsymbol{LFI:white:noise:sensitivity:LFI28:Rad:S:units}{285.3\muKs}

\setsymbol{LFI:white:noise:sensitivity:LFI18:Rad:M}{512.0}
\setsymbol{LFI:white:noise:sensitivity:LFI19:Rad:M}{581.4}
\setsymbol{LFI:white:noise:sensitivity:LFI20:Rad:M}{590.8}
\setsymbol{LFI:white:noise:sensitivity:LFI21:Rad:M}{455.2}
\setsymbol{LFI:white:noise:sensitivity:LFI22:Rad:M}{492.0}
\setsymbol{LFI:white:noise:sensitivity:LFI23:Rad:M}{507.7}
\setsymbol{LFI:white:noise:sensitivity:LFI24:Rad:M}{462.2}
\setsymbol{LFI:white:noise:sensitivity:LFI25:Rad:M}{413.6}
\setsymbol{LFI:white:noise:sensitivity:LFI26:Rad:M}{478.6}
\setsymbol{LFI:white:noise:sensitivity:LFI27:Rad:M}{277.7}
\setsymbol{LFI:white:noise:sensitivity:LFI28:Rad:M}{312.3}
\setsymbol{LFI:white:noise:sensitivity:LFI18:Rad:S}{465.7}
\setsymbol{LFI:white:noise:sensitivity:LFI19:Rad:S}{555.6}
\setsymbol{LFI:white:noise:sensitivity:LFI20:Rad:S}{623.2}
\setsymbol{LFI:white:noise:sensitivity:LFI21:Rad:S}{564.1}
\setsymbol{LFI:white:noise:sensitivity:LFI22:Rad:S}{534.4}
\setsymbol{LFI:white:noise:sensitivity:LFI23:Rad:S}{542.4}
\setsymbol{LFI:white:noise:sensitivity:LFI24:Rad:S}{399.2}
\setsymbol{LFI:white:noise:sensitivity:LFI25:Rad:S}{392.6}
\setsymbol{LFI:white:noise:sensitivity:LFI26:Rad:S}{418.6}
\setsymbol{LFI:white:noise:sensitivity:LFI27:Rad:S}{302.9}
\setsymbol{LFI:white:noise:sensitivity:LFI28:Rad:S}{285.3}

\setsymbol{LFI:knee:frequency:70GHz:units}{29.5\mHz}
\setsymbol{LFI:knee:frequency:44GHz:units}{56.2\mHz}
\setsymbol{LFI:knee:frequency:30GHz:units}{113.7\mHz}

\setsymbol{LFI:knee:frequency:70GHz}{29.5}
\setsymbol{LFI:knee:frequency:44GHz}{56.2}
\setsymbol{LFI:knee:frequency:30GHz}{113.7}

\setsymbol{LFI:knee:frequency:LFI18:Rad:M:units}{16.3\mHz}
\setsymbol{LFI:knee:frequency:LFI19:Rad:M:units}{15.1\mHz}
\setsymbol{LFI:knee:frequency:LFI20:Rad:M:units}{18.7\mHz}
\setsymbol{LFI:knee:frequency:LFI21:Rad:M:units}{37.2\mHz}
\setsymbol{LFI:knee:frequency:LFI22:Rad:M:units}{12.7\mHz}
\setsymbol{LFI:knee:frequency:LFI23:Rad:M:units}{34.6\mHz}
\setsymbol{LFI:knee:frequency:LFI24:Rad:M:units}{46.2\mHz}
\setsymbol{LFI:knee:frequency:LFI25:Rad:M:units}{24.9\mHz}
\setsymbol{LFI:knee:frequency:LFI26:Rad:M:units}{67.6\mHz}
\setsymbol{LFI:knee:frequency:LFI27:Rad:M:units}{187.4\mHz}
\setsymbol{LFI:knee:frequency:LFI28:Rad:M:units}{122.2\mHz}
\setsymbol{LFI:knee:frequency:LFI18:Rad:S:units}{17.7\mHz}
\setsymbol{LFI:knee:frequency:LFI19:Rad:S:units}{22.0\mHz}
\setsymbol{LFI:knee:frequency:LFI20:Rad:S:units}{8.7\mHz}
\setsymbol{LFI:knee:frequency:LFI21:Rad:S:units}{25.9\mHz}
\setsymbol{LFI:knee:frequency:LFI22:Rad:S:units}{15.8\mHz}
\setsymbol{LFI:knee:frequency:LFI23:Rad:S:units}{129.8\mHz}
\setsymbol{LFI:knee:frequency:LFI24:Rad:S:units}{100.9\mHz}
\setsymbol{LFI:knee:frequency:LFI25:Rad:S:units}{38.9\mHz}
\setsymbol{LFI:knee:frequency:LFI26:Rad:S:units}{58.9\mHz}
\setsymbol{LFI:knee:frequency:LFI27:Rad:S:units}{104.4\mHz}
\setsymbol{LFI:knee:frequency:LFI28:Rad:S:units}{40.7\mHz}

\setsymbol{LFI:knee:frequency:LFI18:Rad:M}{16.3}
\setsymbol{LFI:knee:frequency:LFI19:Rad:M}{15.1}
\setsymbol{LFI:knee:frequency:LFI20:Rad:M}{18.7}
\setsymbol{LFI:knee:frequency:LFI21:Rad:M}{37.2}
\setsymbol{LFI:knee:frequency:LFI22:Rad:M}{12.7}
\setsymbol{LFI:knee:frequency:LFI23:Rad:M}{34.6}
\setsymbol{LFI:knee:frequency:LFI24:Rad:M}{46.2}
\setsymbol{LFI:knee:frequency:LFI25:Rad:M}{24.9}
\setsymbol{LFI:knee:frequency:LFI26:Rad:M}{67.6}
\setsymbol{LFI:knee:frequency:LFI27:Rad:M}{187.4}
\setsymbol{LFI:knee:frequency:LFI28:Rad:M}{122.2}
\setsymbol{LFI:knee:frequency:LFI18:Rad:S}{17.7}
\setsymbol{LFI:knee:frequency:LFI19:Rad:S}{22.0}
\setsymbol{LFI:knee:frequency:LFI20:Rad:S}{8.7}
\setsymbol{LFI:knee:frequency:LFI21:Rad:S}{25.9}
\setsymbol{LFI:knee:frequency:LFI22:Rad:S}{15.8}
\setsymbol{LFI:knee:frequency:LFI23:Rad:S}{129.8}
\setsymbol{LFI:knee:frequency:LFI24:Rad:S}{100.9}
\setsymbol{LFI:knee:frequency:LFI25:Rad:S}{38.9}
\setsymbol{LFI:knee:frequency:LFI26:Rad:S}{58.9}
\setsymbol{LFI:knee:frequency:LFI27:Rad:S}{104.4}
\setsymbol{LFI:knee:frequency:LFI28:Rad:S}{40.7}

\setsymbol{LFI:slope:70GHz:units}{$-1.03$\mHz}
\setsymbol{LFI:slope:44GHz:units}{$-0.89$\mHz}
\setsymbol{LFI:slope:30GHz:units}{$-0.87$\mHz}

\setsymbol{LFI:slope:70GHz}{$-1.03$}
\setsymbol{LFI:slope:44GHz}{$-0.89$}
\setsymbol{LFI:slope:30GHz}{$-0.87$}

\setsymbol{LFI:slope:LFI18:Rad:M:units}{$-1.04$\mHz}
\setsymbol{LFI:slope:LFI19:Rad:M:units}{$-1.09$\mHz}
\setsymbol{LFI:slope:LFI20:Rad:M:units}{$-0.69$\mHz}
\setsymbol{LFI:slope:LFI21:Rad:M:units}{$-1.56$\mHz}
\setsymbol{LFI:slope:LFI22:Rad:M:units}{$-1.01$\mHz}
\setsymbol{LFI:slope:LFI23:Rad:M:units}{$-0.96$\mHz}
\setsymbol{LFI:slope:LFI24:Rad:M:units}{$-0.83$\mHz}
\setsymbol{LFI:slope:LFI25:Rad:M:units}{$-0.91$\mHz}
\setsymbol{LFI:slope:LFI26:Rad:M:units}{$-0.95$\mHz}
\setsymbol{LFI:slope:LFI27:Rad:M:units}{$-0.87$\mHz}
\setsymbol{LFI:slope:LFI28:Rad:M:units}{$-0.88$\mHz}
\setsymbol{LFI:slope:LFI18:Rad:S:units}{$-1.15$\mHz}
\setsymbol{LFI:slope:LFI19:Rad:S:units}{$-1.00$\mHz}
\setsymbol{LFI:slope:LFI20:Rad:S:units}{$-0.95$\mHz}
\setsymbol{LFI:slope:LFI21:Rad:S:units}{$-0.92$\mHz}
\setsymbol{LFI:slope:LFI22:Rad:S:units}{$-1.01$\mHz}
\setsymbol{LFI:slope:LFI23:Rad:S:units}{$-0.95$\mHz}
\setsymbol{LFI:slope:LFI24:Rad:S:units}{$-0.73$\mHz}
\setsymbol{LFI:slope:LFI25:Rad:S:units}{$-1.16$\mHz}
\setsymbol{LFI:slope:LFI26:Rad:S:units}{$-0.79$\mHz}
\setsymbol{LFI:slope:LFI27:Rad:S:units}{$-0.82$\mHz}
\setsymbol{LFI:slope:LFI28:Rad:S:units}{$-0.91$\mHz}

\setsymbol{LFI:slope:LFI18:Rad:M}{$-1.04$}
\setsymbol{LFI:slope:LFI19:Rad:M}{$-1.09$}
\setsymbol{LFI:slope:LFI20:Rad:M}{$-0.69$}
\setsymbol{LFI:slope:LFI21:Rad:M}{$-1.56$}
\setsymbol{LFI:slope:LFI22:Rad:M}{$-1.01$}
\setsymbol{LFI:slope:LFI23:Rad:M}{$-0.96$}
\setsymbol{LFI:slope:LFI24:Rad:M}{$-0.83$}
\setsymbol{LFI:slope:LFI25:Rad:M}{$-0.91$}
\setsymbol{LFI:slope:LFI26:Rad:M}{$-0.95$}
\setsymbol{LFI:slope:LFI27:Rad:M}{$-0.87$}
\setsymbol{LFI:slope:LFI28:Rad:M}{$-0.88$}
\setsymbol{LFI:slope:LFI18:Rad:S}{$-1.15$}
\setsymbol{LFI:slope:LFI19:Rad:S}{$-1.00$}
\setsymbol{LFI:slope:LFI20:Rad:S}{$-0.95$}
\setsymbol{LFI:slope:LFI21:Rad:S}{$-0.92$}
\setsymbol{LFI:slope:LFI22:Rad:S}{$-1.01$}
\setsymbol{LFI:slope:LFI23:Rad:S}{$-0.95$}
\setsymbol{LFI:slope:LFI24:Rad:S}{$-0.73$}
\setsymbol{LFI:slope:LFI25:Rad:S}{$-1.16$}
\setsymbol{LFI:slope:LFI26:Rad:S}{$-0.79$}
\setsymbol{LFI:slope:LFI27:Rad:S}{$-0.82$}
\setsymbol{LFI:slope:LFI28:Rad:S}{$-0.91$}

\setsymbol{LFI:FWHM:70GHz:units}{13\parcm01}
\setsymbol{LFI:FWHM:44GHz:units}{27\parcm92}
\setsymbol{LFI:FWHM:30GHz:units}{32\parcm65}

\setsymbol{LFI:FWHM:70GHz}{13.01}
\setsymbol{LFI:FWHM:44GHz}{27.92}
\setsymbol{LFI:FWHM:30GHz}{32.65}

\setsymbol{LFI:FWHM:LFI18:units}{13\parcm39}
\setsymbol{LFI:FWHM:LFI19:units}{13\parcm01}
\setsymbol{LFI:FWHM:LFI20:units}{12\parcm75}
\setsymbol{LFI:FWHM:LFI21:units}{12\parcm74}
\setsymbol{LFI:FWHM:LFI22:units}{12\parcm87}
\setsymbol{LFI:FWHM:LFI23:units}{13\parcm27}
\setsymbol{LFI:FWHM:LFI24:units}{22\parcm98}
\setsymbol{LFI:FWHM:LFI25:units}{30\parcm46}
\setsymbol{LFI:FWHM:LFI26:units}{30\parcm31}
\setsymbol{LFI:FWHM:LFI27:units}{32\parcm65}
\setsymbol{LFI:FWHM:LFI28:units}{32\parcm66}

\setsymbol{LFI:FWHM:LFI18}{13.39}
\setsymbol{LFI:FWHM:LFI19}{13.01}
\setsymbol{LFI:FWHM:LFI20}{12.75}
\setsymbol{LFI:FWHM:LFI21}{12.74}
\setsymbol{LFI:FWHM:LFI22}{12.87}
\setsymbol{LFI:FWHM:LFI23}{13.27}
\setsymbol{LFI:FWHM:LFI24}{22.98}
\setsymbol{LFI:FWHM:LFI25}{30.46}
\setsymbol{LFI:FWHM:LFI26}{30.31}
\setsymbol{LFI:FWHM:LFI27}{32.65}
\setsymbol{LFI:FWHM:LFI28}{32.66}

\setsymbol{LFI:FWHM:uncertainty:LFI18:units}{0.170\arcm}
\setsymbol{LFI:FWHM:uncertainty:LFI19:units}{0.174\arcm}
\setsymbol{LFI:FWHM:uncertainty:LFI20:units}{0.170\arcm}
\setsymbol{LFI:FWHM:uncertainty:LFI21:units}{0.156\arcm}
\setsymbol{LFI:FWHM:uncertainty:LFI22:units}{0.164\arcm}
\setsymbol{LFI:FWHM:uncertainty:LFI23:units}{0.171\arcm}
\setsymbol{LFI:FWHM:uncertainty:LFI24:units}{0.652\arcm}
\setsymbol{LFI:FWHM:uncertainty:LFI25:units}{1.075\arcm}
\setsymbol{LFI:FWHM:uncertainty:LFI26:units}{1.131\arcm}
\setsymbol{LFI:FWHM:uncertainty:LFI27:units}{1.266\arcm}
\setsymbol{LFI:FWHM:uncertainty:LFI28:units}{1.287\arcm}

\setsymbol{LFI:FWHM:uncertainty:LFI18}{0.170}
\setsymbol{LFI:FWHM:uncertainty:LFI19}{0.174}
\setsymbol{LFI:FWHM:uncertainty:LFI20}{0.170}
\setsymbol{LFI:FWHM:uncertainty:LFI21}{0.156}
\setsymbol{LFI:FWHM:uncertainty:LFI22}{0.164}
\setsymbol{LFI:FWHM:uncertainty:LFI23}{0.171}
\setsymbol{LFI:FWHM:uncertainty:LFI24}{0.652}
\setsymbol{LFI:FWHM:uncertainty:LFI25}{1.075}
\setsymbol{LFI:FWHM:uncertainty:LFI26}{1.131}
\setsymbol{LFI:FWHM:uncertainty:LFI27}{1.266}
\setsymbol{LFI:FWHM:uncertainty:LFI28}{1.287}

\setsymbol{HFI:center:frequency:100GHz:units}{100\,GHz}
\setsymbol{HFI:center:frequency:143GHz:units}{143\,GHz}
\setsymbol{HFI:center:frequency:217GHz:units}{217\,GHz}
\setsymbol{HFI:center:frequency:353GHz:units}{353\,GHz}
\setsymbol{HFI:center:frequency:545GHz:units}{545\,GHz}
\setsymbol{HFI:center:frequency:857GHz:units}{857\,GHz}

\setsymbol{HFI:center:frequency:100GHz}{100}
\setsymbol{HFI:center:frequency:143GHz}{143}
\setsymbol{HFI:center:frequency:217GHz}{217}
\setsymbol{HFI:center:frequency:353GHz}{353}
\setsymbol{HFI:center:frequency:545GHz}{545}
\setsymbol{HFI:center:frequency:857GHz}{857}

\setsymbol{HFI:Ndetectors:100GHz}{8}
\setsymbol{HFI:Ndetectors:143GHz}{11}
\setsymbol{HFI:Ndetectors:217GHz}{12}
\setsymbol{HFI:Ndetectors:353GHz}{12}
\setsymbol{HFI:Ndetectors:545GHz}{3}
\setsymbol{HFI:Ndetectors:857GHz}{4}

\setsymbol{HFI:FWHM:Maps:100GHz:units}{9\parcm88}
\setsymbol{HFI:FWHM:Maps:143GHz:units}{7\parcm18}
\setsymbol{HFI:FWHM:Maps:217GHz:units}{4\parcm87}
\setsymbol{HFI:FWHM:Maps:353GHz:units}{4\parcm65}
\setsymbol{HFI:FWHM:Maps:545GHz:units}{4\parcm72}
\setsymbol{HFI:FWHM:Maps:857GHz:units}{4\parcm39}
\setsymbol{HFI:FWHM:Maps:100GHz}{9.88}
\setsymbol{HFI:FWHM:Maps:143GHz}{7.18}
\setsymbol{HFI:FWHM:Maps:217GHz}{4.87}
\setsymbol{HFI:FWHM:Maps:353GHz}{4.65}
\setsymbol{HFI:FWHM:Maps:545GHz}{4.72}
\setsymbol{HFI:FWHM:Maps:857GHz}{4.39}

\setsymbol{HFI:beam:ellipticity:Maps:100GHz}{1.15}
\setsymbol{HFI:beam:ellipticity:Maps:143GHz}{1.01}
\setsymbol{HFI:beam:ellipticity:Maps:217GHz}{1.06}
\setsymbol{HFI:beam:ellipticity:Maps:353GHz}{1.05}
\setsymbol{HFI:beam:ellipticity:Maps:545GHz}{1.14}
\setsymbol{HFI:beam:ellipticity:Maps:857GHz}{1.19}

\setsymbol{HFI:FWHM:Mars:100GHz:units}{9\parcm37}
\setsymbol{HFI:FWHM:Mars:143GHz:units}{7\parcm04}
\setsymbol{HFI:FWHM:Mars:217GHz:units}{4\parcm68}
\setsymbol{HFI:FWHM:Mars:353GHz:units}{4\parcm43}
\setsymbol{HFI:FWHM:Mars:545GHz:units}{3\parcm80}
\setsymbol{HFI:FWHM:Mars:857GHz:units}{3\parcm67}

\setsymbol{HFI:FWHM:Mars:100GHz}{9.37}
\setsymbol{HFI:FWHM:Mars:143GHz}{7.04}
\setsymbol{HFI:FWHM:Mars:217GHz}{4.68}
\setsymbol{HFI:FWHM:Mars:353GHz}{4.43}
\setsymbol{HFI:FWHM:Mars:545GHz}{3.80}
\setsymbol{HFI:FWHM:Mars:857GHz}{3.67}

\setsymbol{HFI:beam:ellipticity:Mars:100GHz}{1.18}
\setsymbol{HFI:beam:ellipticity:Mars:143GHz}{1.03}
\setsymbol{HFI:beam:ellipticity:Mars:217GHz}{1.14}
\setsymbol{HFI:beam:ellipticity:Mars:353GHz}{1.09}
\setsymbol{HFI:beam:ellipticity:Mars:545GHz}{1.25}
\setsymbol{HFI:beam:ellipticity:Mars:857GHz}{1.03}

\setsymbol{HFI:CMB:relative:calibration:100GHz}{$\lsim 1\%$}
\setsymbol{HFI:CMB:relative:calibration:143GHz}{$\lsim 1\%$}
\setsymbol{HFI:CMB:relative:calibration:217GHz}{$\lsim 1\%$}
\setsymbol{HFI:CMB:relative:calibration:353GHz}{$\lsim 1\%$}
\setsymbol{HFI:CMB:relative:calibration:545GHz}{}
\setsymbol{HFI:CMB:relative:calibration:857GHz}{}

\setsymbol{HFI:CMB:absolute:calibration:100GHz}{$\lsim 2\%$}
\setsymbol{HFI:CMB:absolute:calibration:143GHz}{$\lsim 2\%$}
\setsymbol{HFI:CMB:absolute:calibration:217GHz}{$\lsim 2\%$}
\setsymbol{HFI:CMB:absolute:calibration:353GHz}{$\lsim 2\%$}
\setsymbol{HFI:CMB:absolute:calibration:545GHz}{}
\setsymbol{HFI:CMB:absolute:calibration:857GHz}{}

\setsymbol{HFI:FIRAS:gain:calibration:accuracy:statistical:100GHz}{}
\setsymbol{HFI:FIRAS:gain:calibration:accuracy:statistical:143GHz}{}
\setsymbol{HFI:FIRAS:gain:calibration:accuracy:statistical:217GHz}{}
\setsymbol{HFI:FIRAS:gain:calibration:accuracy:statistical:353GHz}{2.5\%}
\setsymbol{HFI:FIRAS:gain:calibration:accuracy:statistical:545GHz}{1\%}
\setsymbol{HFI:FIRAS:gain:calibration:accuracy:statistical:857GHz}{0.5\%}

\setsymbol{HFI:FIRAS:gain:calibration:accuracy:systematic:100GHz}{}
\setsymbol{HFI:FIRAS:gain:calibration:accuracy:systematic:143GHz}{}
\setsymbol{HFI:FIRAS:gain:calibration:accuracy:systematic:217GHz}{}
\setsymbol{HFI:FIRAS:gain:calibration:accuracy:systematic:353GHz}{}
\setsymbol{HFI:FIRAS:gain:calibration:accuracy:systematic:545GHz}{7\%}
\setsymbol{HFI:FIRAS:gain:calibration:accuracy:systematic:857GHz}{7\%}

\setsymbol{HFI:FIRAS:zero:point:accuracy:100GHz:units}{0.8\MJysr}
\setsymbol{HFI:FIRAS:zero:point:accuracy:143GHz:units}{}
\setsymbol{HFI:FIRAS:zero:point:accuracy:217GHz:units}{}
\setsymbol{HFI:FIRAS:zero:point:accuracy:353GHz:units}{1.4\MJysr}
\setsymbol{HFI:FIRAS:zero:point:accuracy:545GHz:units}{2.2\MJysr}
\setsymbol{HFI:FIRAS:zero:point:accuracy:857GHz:units}{1.7\MJysr}

\setsymbol{HFI:FIRAS:zero:point:accuracy:100GHz}{0.8}
\setsymbol{HFI:FIRAS:zero:point:accuracy:143GHz}{}
\setsymbol{HFI:FIRAS:zero:point:accuracy:217GHz}{}
\setsymbol{HFI:FIRAS:zero:point:accuracy:353GHz}{1.4}
\setsymbol{HFI:FIRAS:zero:point:accuracy:545GHz}{2.2}
\setsymbol{HFI:FIRAS:zero:point:accuracy:857GHz}{1.7}

\setsymbol{HFI:unit:conversion:100GHz:units}{0.2415\MJysrmK}
\setsymbol{HFI:unit:conversion:143GHz:units}{0.3694\MJysrmK}
\setsymbol{HFI:unit:conversion:217GHz:units}{0.4811\MJysrmK}
\setsymbol{HFI:unit:conversion:353GHz:units}{0.2883\MJysrmK}
\setsymbol{HFI:unit:conversion:545GHz:units}{0.05826\MJysrmK}
\setsymbol{HFI:unit:conversion:857GHz:units}{0.002238\MJysrmK}

\setsymbol{HFI:unit:conversion:100GHz}{0.2415}
\setsymbol{HFI:unit:conversion:143GHz}{0.3694}
\setsymbol{HFI:unit:conversion:217GHz}{0.4811}
\setsymbol{HFI:unit:conversion:353GHz}{0.2883}
\setsymbol{HFI:unit:conversion:545GHz}{0.05826}
\setsymbol{HFI:unit:conversion:857GHz}{0.002238}

\setsymbol{HFI:colour:correction:alpha=-2:V1.01:100GHz}{0.9893}
\setsymbol{HFI:colour:correction:alpha=-2:V1.01:143GHz}{0.9759}
\setsymbol{HFI:colour:correction:alpha=-2:V1.01:217GHz}{1.0007}
\setsymbol{HFI:colour:correction:alpha=-2:V1.01:353GHz}{1.0028}
\setsymbol{HFI:colour:correction:alpha=-2:V1.01:545GHz}{1.0019}
\setsymbol{HFI:colour:correction:alpha=-2:V1.01:857GHz}{0.9889}

\setsymbol{HFI:colour:correction:alpha=0:V1.01:100GHz}{1.0008}
\setsymbol{HFI:colour:correction:alpha=0:V1.01:143GHz}{1.0148}
\setsymbol{HFI:colour:correction:alpha=0:V1.01:217GHz}{0.9909}
\setsymbol{HFI:colour:correction:alpha=0:V1.01:353GHz}{0.9888}
\setsymbol{HFI:colour:correction:alpha=0:V1.01:545GHz}{0.9878}
\setsymbol{HFI:colour:correction:alpha=0:V1.01:857GHz}{1.0014}

\def\xmm{{\it XMM-Newton}}

\newfont{\gwpfont}{cmssq8 scaled 1000}
\newcommand{\rexcess}{{\gwpfont REXCESS}}

\def\xmmesz{{\rm ESZ}--{\it XMM\/}}

\def\M500{M_{500}}
\def\R500{R_{500}}
\def\L500{L_{500}}

\def\YX {Y_{\rm X}}

\def\T500{\theta_{\rm 500}}
\def\Y500{\theta_{\rm 500}}

\def\Mv {M_{\rm 500}}
\def \Rv {R_{500}}
\def\keV {\rm keV}
\def\Yv {Y_{500}}

\def\msol {{\rm M_{\odot}}}

\def\lesssim{\mathrel{\hbox{\rlap{\hbox{\lower4pt\hbox{$\sim$}}}\hbox{$<$}}}}
\def\gtrsim{\mathrel{\hbox{\rlap{\hbox{\lower4pt\hbox{$\sim$}}}\hbox{$>$}}}}

\newcommand{\propsim}{\lower 3pt \hbox{$\, \buildrel {\textstyle
       \propto}\over {\textstyle \sim}\,$}}

\begin{document}
\author{\small
Planck Collaboration:
P.~A.~R.~Ade\inst{88}
\and
N.~Aghanim\inst{60}
\and
M.~Arnaud\inst{75}
\and
M.~Ashdown\inst{72, 6}
\and
F.~Atrio-Barandela\inst{19}
\and
J.~Aumont\inst{60}
\and
C.~Baccigalupi\inst{87}
\and
A.~Balbi\inst{38}
\and
A.~J.~Banday\inst{96, 9}
\and
R.~B.~Barreiro\inst{69}
% ...
}
\institute{\small
APC, AstroParticule et Cosmologie, Universit\'{e} Paris Diderot, CNRS/IN2P3, CEA/lrfu, Observatoire de Paris, Sorbonne Paris Cit\'{e}, 10, rue Alice Domon et L\'{e}onie Duquet, 75205 Paris Cedex 13, France\\
\and
Aalto University Mets\"{a}hovi Radio Observatory, Mets\"{a}hovintie 114, FIN-02540 Kylm\"{a}l\"{a}, Finland\\
\and
Academy of Sciences of Tatarstan, Bauman Str., 20, Kazan, 420111, Republic of Tatarstan, Russia\\
\and
Agenzia Spaziale Italiana Science Data Center, c/o ESRIN, via Galileo Galilei, Frascati, Italy\\
\and
Agenzia Spaziale Italiana, Viale Liegi 26, Roma, Italy\\
\and
Astrophysics Group, Cavendish Laboratory, University of Cambridge, J J Thomson Avenue, Cambridge CB3 0HE, U.K.\\
\and
Atacama Large Millimeter/submillimeter Array, ALMA Santiago Central Offices, Alonso de Cordova 3107, Vitacura, Casilla 763 0355, Santiago, Chile\\
\and
CITA, University of Toronto, 60 St. George St., Toronto, ON M5S 3H8, Canada\\
\and
CNRS, IRAP, 9 Av. colonel Roche, BP 44346, F-31028 Toulouse cedex 4, France\\
\and
California Institute of Technology, Pasadena, California, U.S.A.\\
\and
Centre of Mathematics for Applications, University of Oslo, Blindern, Oslo, Norway\\
\and
Centro de Astrof\'{\i}sica, Universidade do Porto, Rua das Estrelas, 4150-762 Porto, Portugal\\
\and
Centro de Estudios de F\'{i}sica del Cosmos de Arag\'{o}n (CEFCA), Plaza San Juan, 1, planta 2, E-44001, Teruel, Spain\\
\and
Computational Cosmology Center, Lawrence Berkeley National Laboratory, Berkeley, California, U.S.A.\\
\and
Consejo Superior de Investigaciones Cient\'{\i}ficas (CSIC), Madrid, Spain\\
\and
DSM/Irfu/SPP, CEA-Saclay, F-91191 Gif-sur-Yvette Cedex, France\\
\and
DTU Space, National Space Institute, Technical University of Denmark, Elektrovej 327, DK-2800 Kgs. Lyngby, Denmark\\
\and
D\'{e}partement de Physique Th\'{e}orique, Universit\'{e} de Gen\`{e}ve, 24, Quai E. Ansermet,1211 Gen\`{e}ve 4, Switzerland\\
\and
Departamento de F\'{\i}sica Fundamental, Facultad de Ciencias, Universidad de Salamanca, 37008 Salamanca, Spain\\
\and
Departamento de F\'{\i}sica, Universidad de Oviedo, Avda. Calvo Sotelo s/n, Oviedo, Spain\\
\and
Department of Astronomy and Geodesy, Kazan Federal University,  Kremlevskaya Str., 18, Kazan, 420008, Russia\\
\and
Department of Astrophysics, IMAPP, Radboud University, P.O. Box 9010, 6500 GL Nijmegen,  The Netherlands\\
\and
Department of Physics \& Astronomy, University of British Columbia, 6224 Agricultural Road, Vancouver, British Columbia, Canada\\
\and
Department of Physics and Astronomy, Dana and David Dornsife College of Letter, Arts and Sciences, University of Southern California, Los Angeles, CA 90089, U.S.A.\\
\and
Department of Physics and Astronomy, University of Sussex, Brighton BN1 9QH, U.K.\\
\and
Department of Physics, Gustaf H\"{a}llstr\"{o}min katu 2a, University of Helsinki, Helsinki, Finland\\
\and
Department of Physics, Princeton University, Princeton, New Jersey, U.S.A.\\
\and
Department of Physics, University of California, Berkeley, California, U.S.A.\\
\and
Department of Physics, University of California, One Shields Avenue, Davis, California, U.S.A.\\
\and
Department of Physics, University of California, Santa Barbara, California, U.S.A.\\
\and
Department of Physics, University of Illinois at Urbana-Champaign, 1110 West Green Street, Urbana, Illinois, U.S.A.\\
\and
Department of Statistics, Purdue University, 250 N. University Street, West Lafayette, Indiana, U.S.A.\\
\and
Dipartimento di Fisica e Astronomia G. Galilei, Universit\`{a} degli Studi di Padova, via Marzolo 8, 35131 Padova, Italy\\
\and
Dipartimento di Fisica, Universit\`{a} La Sapienza, P. le A. Moro 2, Roma, Italy\\
\and
Dipartimento di Fisica, Universit\`{a} degli Studi di Milano, Via Celoria, 16, Milano, Italy\\
\and
Dipartimento di Fisica, Universit\`{a} degli Studi di Trieste, via A. Valerio 2, Trieste, Italy\\
\and
Dipartimento di Fisica, Universit\`{a} di Ferrara, Via Saragat 1, 44122 Ferrara, Italy\\
\and
Dipartimento di Fisica, Universit\`{a} di Roma Tor Vergata, Via della Ricerca Scientifica, 1, Roma, Italy\\
\and
Dipartimento di Matematica, Universit\`{a} di Roma Tor Vergata, Via della Ricerca Scientifica, 1, Roma, Italy\\
\and
Discovery Center, Niels Bohr Institute, Blegdamsvej 17, Copenhagen, Denmark\\
\and
Dpto. Astrof\'{i}sica, Universidad de La Laguna (ULL), E-38206 La Laguna, Tenerife, Spain\\
\and
European Southern Observatory, ESO Vitacura, Alonso de Cordova 3107, Vitacura, Casilla 19001, Santiago, Chile\\
\and
European Space Agency, ESAC, Planck Science Office, Camino bajo del Castillo, s/n, Urbanizaci\'{o}n Villafranca del Castillo, Villanueva de la Ca\~{n}ada, Madrid, Spain\\
\and
European Space Agency, ESTEC, Keplerlaan 1, 2201 AZ Noordwijk, The Netherlands\\
\and
GEPI, Observatoire de Paris, Section de Meudon, 5 Place J. Janssen, 92195 Meudon Cedex, France\\
\and
Helsinki Institute of Physics, Gustaf H\"{a}llstr\"{o}min katu 2, University of Helsinki, Helsinki, Finland\\
\and
INAF - Osservatorio Astronomico di Padova, Vicolo dell'Osservatorio 5, Padova, Italy\\
\and
INAF - Osservatorio Astronomico di Roma, via di Frascati 33, Monte Porzio Catone, Italy\\
\and
INAF - Osservatorio Astronomico di Trieste, Via G.B. Tiepolo 11, Trieste, Italy\\
\and
INAF Istituto di Radioastronomia, Via P. Gobetti 101, 40129 Bologna, Italy\\
\and
INAF/IASF Bologna, Via Gobetti 101, Bologna, Italy\\
\and
INAF/IASF Milano, Via E. Bassini 15, Milano, Italy\\
\and
INFN, Sezione di Roma 1, Universit`{a} di Roma Sapienza, Piazzale Aldo Moro 2, 00185, Roma, Italy\\
\and
IPAG: Institut de Plan\'{e}tologie et d'Astrophysique de Grenoble, Universit\'{e} Joseph Fourier, Grenoble 1 / CNRS-INSU, UMR 5274, Grenoble, F-38041, France\\
\and
IUCAA, Post Bag 4, Ganeshkhind, Pune University Campus, Pune 411 007, India\\
\and
Imperial College London, Astrophysics group, Blackett Laboratory, Prince Consort Road, London, SW7 2AZ, U.K.\\
\and
Infrared Processing and Analysis Center, California Institute of Technology, Pasadena, CA 91125, U.S.A.\\
\and
Institut N\'{e}el, CNRS, Universit\'{e} Joseph Fourier Grenoble I, 25 rue des Martyrs, Grenoble, France\\
\and
Institut Universitaire de France, 103, bd Saint-Michel, 75005, Paris, France\\
\and
Institut d'Astrophysique Spatiale, CNRS (UMR8617) Universit\'{e} Paris-Sud 11, B\^{a}timent 121, Orsay, France\\
\and
Institut d'Astrophysique de Paris, CNRS (UMR7095), 98 bis Boulevard Arago, F-75014, Paris, France\\
\and
Institut de Ci\`{e}ncies de l'Espai, CSIC/IEEC, Facultat de Ci\`{e}ncies, Campus UAB, Torre C5 par-2, Bellaterra 08193, Spain\\
\and
Institute for Space Sciences, Bucharest-Magurale, Romania\\
\and
Institute of Astro and Particle Physics, Technikerstrasse 25/8, University of Innsbruck, A-6020, Innsbruck, Austria\\
\and
Institute of Astronomy and Astrophysics, Academia Sinica, Taipei, Taiwan\\
\and
Institute of Astronomy, University of Cambridge, Madingley Road, Cambridge CB3 0HA, U.K.\\
\and
Institute of Theoretical Astrophysics, University of Oslo, Blindern, Oslo, Norway\\
\and
Instituto de Astrof\'{\i}sica de Canarias, C/V\'{\i}a L\'{a}ctea s/n, La Laguna, Tenerife, Spain\\
\and
Instituto de F\'{\i}sica de Cantabria (CSIC-Universidad de Cantabria), Avda. de los Castros s/n, Santander, Spain\\
\and
Jet Propulsion Laboratory, California Institute of Technology, 4800 Oak Grove Drive, Pasadena, California, U.S.A.\\
\and
Jodrell Bank Centre for Astrophysics, Alan Turing Building, School of Physics and Astronomy, The University of Manchester, Oxford Road, Manchester, M13 9PL, U.K.\\
\and
Kavli Institute for Cosmology Cambridge, Madingley Road, Cambridge, CB3 0HA, U.K.\\
\and
LAL, Universit\'{e} Paris-Sud, CNRS/IN2P3, Orsay, France\\
\and
LERMA, CNRS, Observatoire de Paris, 61 Avenue de l'Observatoire, Paris, France\\
\and
Laboratoire AIM, IRFU/Service d'Astrophysique - CEA/DSM - CNRS - Universit\'{e} Paris Diderot, B\^{a}t. 709, CEA-Saclay, F-91191 Gif-sur-Yvette Cedex, France\\
\and
Laboratoire Traitement et Communication de l'Information, CNRS (UMR 5141) and T\'{e}l\'{e}com ParisTech, 46 rue Barrault F-75634 Paris Cedex 13, France\\
\and
Laboratoire de Physique Subatomique et de Cosmologie, Universit\'{e} Joseph Fourier Grenoble I, CNRS/IN2P3, Institut National Polytechnique de Grenoble, 53 rue des Martyrs, 38026 Grenoble cedex, France\\
\and
Laboratoire de Physique Th\'{e}orique, Universit\'{e} Paris-Sud 11 \& CNRS, B\^{a}timent 210, 91405 Orsay, France\\
\and
Lawrence Berkeley National Laboratory, Berkeley, California, U.S.A.\\
\and
Max-Planck-Institut f\"{u}r Astrophysik, Karl-Schwarzschild-Str. 1, 85741 Garching, Germany\\
\and
Max-Planck-Institut f\"{u}r Extraterrestrische Physik, Giessenbachstra{\ss}e, 85748 Garching, Germany\\
\and
MilliLab, VTT Technical Research Centre of Finland, Tietotie 3, Espoo, Finland\\
\and
National University of Ireland, Department of Experimental Physics, Maynooth, Co. Kildare, Ireland\\
\and
Niels Bohr Institute, Blegdamsvej 17, Copenhagen, Denmark\\
\and
Observational Cosmology, Mail Stop 367-17, California Institute of Technology, Pasadena, CA, 91125, U.S.A.\\
\and
Optical Science Laboratory, University College London, Gower Street, London, U.K.\\
\and
SISSA, Astrophysics Sector, via Bonomea 265, 34136, Trieste, Italy\\
\and
School of Physics and Astronomy, Cardiff University, Queens Buildings, The Parade, Cardiff, CF24 3AA, U.K.\\
\and
Space Research Institute (IKI), Profsoyuznaya 84/32, Moscow, Russia\\
\and
Space Research Institute (IKI), Russian Academy of Sciences, Profsoyuznaya Str, 84/32, Moscow, 117997, Russia\\
\and
Space Sciences Laboratory, University of California, Berkeley, California, U.S.A.\\
\and
Stanford University, Dept of Physics, Varian Physics Bldg, 382 Via Pueblo Mall, Stanford, California, U.S.A.\\
\and
T\"{U}B\.{I}TAK National Observatory, Akdeniz University Campus, 07058, Antalya, Turkey\\
\and
UPMC Univ Paris 06, UMR7095, 98 bis Boulevard Arago, F-75014, Paris, France\\
\and
Universit\'{e} Denis Diderot (Paris 7), 75205 Paris Cedex 13, France\\
\and
Universit\'{e} de Toulouse, UPS-OMP, IRAP, F-31028 Toulouse cedex 4, France\\
% ...
}

   \title{\textit{Planck\/} Intermediate Results. V. Pressure profiles of galaxy
 clusters from the Sunyaev-Zeldovich effect}
\authorrunning{Planck Collaboration}
\titlerunning{Pressure profiles of galaxy clusters from the Sunyaev-Zeldovich effect}

   \date{Received 17 July 2012; accepted 29 October 2012}
  \abstract{
Taking advantage of the all-sky coverage and broad frequency range of the
\Planck\ satellite, we study the Sunyaev-Zeldovich (SZ) and pressure profiles
of 62 nearby massive clusters detected at high significance in the 14-month
nominal survey.  Careful reconstruction of the SZ signal indicates that most
clusters are individually detected at least out to $\Rv$.  By stacking the
radial profiles, we have statistically detected the radial SZ signal out to
$3\times \Rv$, i.e., at a density contrast of about 50--100, though the
dispersion about the mean profile dominates the statistical errors across the
whole radial range.
Our measurement is fully consistent with previous \Planck\ results on
integrated SZ fluxes, further strengthening the agreement between SZ and X-ray
measurements inside $\Rv$. 
Correcting for the effects of the \Planck\ beam, we have calculated the
corresponding pressure profiles.  This new constraint from SZ measurements is
consistent with the X-ray constraints from \xmm\ in the region in which the
profiles overlap (i.e., [0.1--1]$\,\Rv$), and is in fairly good agreement
with theoretical predictions within the expected dispersion.  At larger radii
the average pressure profile {is slightly flatter than most predictions from numerical simulations}.
Combining the SZ and X-ray observed profiles into a joint fit to a generalised  pressure
profile gives best-fit parameters $[P_0, c_{\rm 500}, \gamma, \alpha, \beta]
= [6.41, 1.81, 0.31, 1.33, 4.13]$.
Using a reasonable hypothesis for the gas temperature in the cluster outskirts
we reconstruct from our stacked pressure profile the gas mass fraction profile
out to 3$\,\Rv$. Within the temperature driven uncertainties, our \Planck\
constraints are compatible with the cosmic baryon fraction and expected gas
fraction in halos.}

   \keywords{Cosmology: observations --
                Galaxies: clusters: general -- 
                Galaxies: clusters: intracluster medium --
                Submillimeter: general -- 
                X-rays: general
                }


 \maketitle
%
\section {Introduction}
\label{s:intro}

In a pure hierarchical gravitational collapse scenario in the concordance
$\Lambda$CDM cosmology, concentrations of matter (``halos'') are fully
characterised by their redshift and mean matter density, which in turn are
related to the power spectrum of initial density fluctuations
\citep{Peebles1980}. The scale-free dark matter (DM) collapse drives the
evolution of halo concentration across cosmic time
\citep[see e.g., recent work by][]{bhattacharya11,gao12}, and the ensuing
similarity yields both a universal dark matter distribution and simple global
scaling relations that should describe the entire halo population
\citep{ber85,kaiser95,navarro95,evrard96,navarro97,voi05,arn05}.
However, the observable properties of clusters are determined by the visible
baryonic component, which is subject to more complex physical processes
related to galaxy formation and feedback. As the main baryonic reservoir in
massive halos, the hot gas in the intra-cluster medium (ICM) is the natural
target for studying the physical processes at play and their link to the
underlying cluster DM content. Modelling and understanding the baryon physics
and disentangling the effect of various feedback processes is a very
challenging task \citep[see e.g.,][for a review]{borgani11}, underlining the
need for appropriate observational constraints.  

The ICM attains X-ray emitting temperatures due to gravitational heating
within the halo potential well. X-ray emission is proportional to the square
of the gas density, thus it probes  the denser regions of the hot gas.  The
thermal Sunyaev-Zeldovich (SZ) effect \citep{sunyaev72}, due to inverse Compton
scattering of Cosmic Microwave Background (CMB) photons by the ICM, is
proportional to the thermal gas pressure integrated along the line of sight.
It is sensitive to density and/or temperature variations, such as shocks and
compression. These two independent observational probes are thus complementary
and allow us to further constrain the physics of the ICM. 

Recent X-ray observations based primarily on representative samples have
returned a consistent picture of the scaling and structural properties of
halos, from high mass clusters down to the low mass group regime
\citep[see, e.g.,][]{boe07,vik06,cro08,pra09,arn10,sun11,sun12}.
%
In parallel, SZ observations with instruments such as the Sunyaev-Zeldovich
Array \citep[SZA,][]{muc07}, the South Pole Telescope \citep[SPT,][]{car11},
the Atacama Cosmology Telescope \citep[ACT,][]{swetz11}, and \Planck\
\citep{tauber2010a}\footnote{\Planck\ (\url{http://www.esa.int/Planck}) is a
project of the European Space Agency (ESA) with instruments provided by two
scientific consortia funded by ESA member 
states (in particular the lead countries France and Italy), with contributions
from NASA (USA) and telescope reflectors provided by a collaboration between
ESA and a scientific consortium led and funded by Denmark.}  have recently
started to deliver on the promise of SZ observations for cluster studies.

%
Building on earlier works on smaller cluster samples \citep{ben04,and11}, 
recent results from \Planck\ have underlined the consistency between the X-ray 
and SZ view of the ICM within $\Rv$\footnote{The quantity $\Rv$ corresponds to 
a total density contrast $\delta=500$, as compared to $\rho_{\rm c}(z)$, the 
critical density of the Universe at the cluster redshift.  It is  linked to the 
mass scale by: $\Mv = (4\pi/3)\,500\,\rho_c(z)\,\Rv^3$.}. These constraints 
were achieved using three different approaches. The first, detailed in 
\citet{planck2011-5.2a}, involved bin-averaging of an X-ray selected sample 
from the Meta-Catalogue of X-ray detected Clusters of galaxies 
\citep[MCXC,][]{pif11}. The second, described in \citet{planck2011-5.2b}, 
concerned the comparison of SZ measurements of 62 local clusters from the Early 
Release Compact Source Catalogue 
\citep[ERCSC,][]{planck2011-1.10,planck2011-5.1a} with good quality \xmm\ 
archive data. Finally \citet{planck2011-5.1b} examined the scaling properties 
of 21 newly-detected \Planck\ clusters confirmed by \xmm. These studies have 
provided  well-constrained scaling relations (e.g., $\Yv-M_{500}$, 
$\Yv-L_{X,500}$) in the local Universe, including the first measurement of 
their intrinsic scatter, to be used as a reference for future evolution and 
cosmology studies.  Other dedicated investigations of smaller samples
\citep[e.g.,][]{and11,sifon12} have 
reached conclusions similar to \Planck\ early results regarding the agreement 
between the X-ray and SZ view of  the ICM. 

%
Beyond global properties, similarity of shape in cluster radial quantities has 
long been shown in X-ray observations 
\citep[e.g.,][]{arn01,poi05,vik06,cro08,arn10}, in the optical 
\citep[e.g.,][]{rines06, wojtak10}, and with weak-lensing 
\citep[e.g.,][]{postman12}. For SZ studies, the radial distribution of the 
thermal ICM pressure in massive halos is of particular interest. Recent X-ray 
observations of \rexcess, a representative sample of nearby objects, have shown 
that the ICM pressure distribution, when scaled appropriately, follows an 
approximately ``universal'' shape \citep[][A10 hereafter]{arn10} up to a radius 
of $\Rv$. The small differences in shape for the pressure profiles between 
relaxed and unrelaxed clusters, especially in the central regions, does not 
seem to have an impact on the integrated thermal content within $\Rv$, or on 
other global properties, as noted by \citet{planck2011-5.2b}.

A10 also compared their X-ray pressure profiles to those predicted from 
numerical simulations. Although limited to $\Rv$, good agreement was seen 
between radial pressure profile observations and predictions from various 
numerical simulations within this radius. However, there are presently few 
observational constraints beyond $\Rv$, and consequently in this region the 
shape of the ``universal pressure profile'' was extrapolated according to the 
predictions from numerical simulations \citep{borgani04,nag07sim,piffaretti08}.

The thermodynamical state of the gas beyond $\Rv$ bears the signature of the 
complex physics taking place in the outer parts of the clusters. 
Characterisation of the gas in cluster outskirts is also necessary to unveil 
the level of thermal pressure in the cluster periphery, thus constraining their 
dynamical state. These constraints are crucial for our understanding of the 
formation and evolution of massive halos.
In the cluster surroundings,  accretion along filaments is three-dimensional 
and non-spherical \citep{tozzi01} and the initial conditions of the accretion 
shock are driven by the thermodynamical state of the (pre-shocked) in-falling 
material \citep{voi02,voi03}. The physical origin of any possible pre-heating 
is still unclear, but it may smooth the continuous accretion of substructures. 
Thus the entropy production at the accretion shock may be boosted 
\citep{voi03b,borgani05}. Feedback may also improve the degree of 
thermalisation of the ICM after the shock \citep[with a residual kinetic energy 
of about 10\% beyond $\Rv$,][]{kay04}.

Observations with X-ray telescopes have only recently started to provide 
insight into the physical properties of the gas in the cluster outskirts beyond 
$\Rv$ \citep{geo09,rei09,urb11}. The steep decline of the X-ray emission with 
cluster-centric radius makes such observations extremely challenging with 
current instruments. A deep X-ray study of the Perseus cluster with
{\it Suzaku\/} by  \citet{sim11} indicated that the gas mass fraction in the cluster outskirts was 
overestimated, possibly due to the effect of gas clumping and the 
non-virialised state of the ICM in these regions, as suggested in some 
numerical simulations \citep{nag07sim}. 

The different sensitivity of the SZ effect to the radial ICM distribution means 
that SZ observations have the potential to contribute greatly to the discussion 
on cluster outskirts. The radial pressure distribution of the first SZ cluster 
samples have recently been presented based on observations
by SPT \citep{pla10}, ACT \citep{seh10} 
and SZA/CARMA \citep{bonamente11}. These studies confirmed that the ICM 
properties as seen by SZ and X-ray observations are consistent at least
out to $\Rv$. 
Beyond this radius the SZ effect offers the interesting possibility to further 
constrain the thermal pressure support. \Planck\ is highly competitive in this 
regard. It is the only SZ experiment with a full sky coverage, able to map even 
nearby clusters to their outermost radii and offering the possibility of an 
in-depth statistical study through the combination of many observations. In
this paper we present constraints on the thermal pressure support derived using
SZ observations from the \Planck\ survey. Following our previous methodology
for scaling relations, we investigate these issues from a statistical point of 
view, working with data from 62 local clusters selected from the \Planck\ ESZ 
sample \citep{planck2011-5.1a}, for which there are good quality \xmm\ archival 
data. 

%
~\\[-0.5em]
The paper is organised as follows. In the next section we describe the \Planck\ 
mission and data, together with the sample of galaxy clusters used in this 
study. In Sect.~\ref{s:form} we recall the basic formalism on the SZ effect and 
the parametrisaton of the cluster pressure profile we use throughout the paper. 
 Sect.~\ref{s:recons} is devoted to a detailed description of the processing 
involved in the reconstruction of the SZ and pressure profiles from \Planck\ 
data. In Sects.~\ref{s:szprof} and \ref{s:pprof} we present the stacked 
profile of our sample and its best analytical representation. We discuss the 
comparison with other observational and theoretical constraints in 
Sect.~\ref{s:dis}, before presenting our conclusions in Sect.~\ref{s:con}.

Throughout the paper we adopt a $\Lambda$CDM cosmology with $H_{0} = 70$~ km\ 
s$^{-1}$\ Mpc$^{-1}$, $\Omega_{\rm M} = 0.3$ and $\Omega_\Lambda= 0.7$ . The 
quantity $E(z)$ is the ratio of the Hubble parameter at redshift $z$ to its 
present value, $H_{0}$, i.e., $E(z)^2=\Omega_\mathrm{M} (1+z)^3 + 
\Omega_\mathrm{\Lambda}$.


\section {Data}
\label{s:data}

\subsection {\Planck\ data}\label{s:plck}
  
\Planck\ \citep{tauber2010a, planck2011-1.1} is the third generation space 
mission to measure the anisotropy of the CMB. 
\Planck\ observes the sky in nine frequency bands covering 30--857\,GHz, with 
high sensitivity and angular resolution from 31\arcm\ to 5\arcm. The Low 
Frequency Instrument \citep[LFI;][]{Mandolesi2010, 
Bersanelli2010,planck2011-1.4} covers the 30, 44, and 70\,GHz bands with 
amplifiers cooled to 20\,\hbox{K}. The High Frequency Instrument (HFI; 
\citealt{Lamarre2010, planck2011-1.5}) covers the 100, 143, 217, 353, 545, and 
857\,GHz bands with bolometers cooled to 0.1\,\hbox{K}.  
Early astrophysics results, based on data taken between 13~August 2009 and 
7~June 2010 \citep{planck2011-1.7, planck2011-1.6}, are given in Planck 
Collaboration VIII--XXVI 2011. Intermediate astrophysics results are now based 
on data taken between 13~August 2009 and 27~November 2010.
~\\

We used the full sky maps in the nine \Planck\ frequency bands provided in {\sc 
HEALPix} \citep{gorski2005} $N_{\rm side}=2048$ full resolution.
An error map associated 
to each frequency band is obtained from the difference of the first half and 
second half of the \Planck\ rings for a given position of the satellite. The 
resulting jack-knife maps are basically free from astrophysical emission, 
whilst being a good representation of the statistical instrumental noise and 
systematic error. We adopted a circular Gaussian as the beam pattern for each 
frequency, as described in \citet{planck2011-1.7} and \citet{planck2011-1.6}.
Uncertainties in flux measurements due to beam corrections, map calibrations 
and uncertainties in bandpasses are expected to be small, as discussed 
extensively in \citet{planck2011-5.1a}, \cite{planck2011-5.1b} and
\cite{planck2011-5.2a}.

\subsection{Cluster sample}\label{s:samp}

The first all-sky coverage by the \Planck\ satellite \citep{planck2011-1.1} has 
allowed the detection of dozens of clusters via their SZ signature on the CMB. 
The Early Release SZ sample \citep[ESZ,][]{planck2011-5.1a} comprises 189 
clusters of galaxies characterised over the sub-millimetre to centimetre 
wavelength range. The sample consists of SZ clusters and candidates detected 
with signal-to-noise ratios (S/N) spanning from 6 to 29 in the first all-sky 
survey.  The sample
was thoroughly validated by \Planck\ internal quality assessment, 
external X-ray and optical data cross-correlations, and a multi-frequency 
follow-up programme for confirmation. The ESZ-cluster sample spans over a 
decade in mass, from 0.9 to 15$\times10^{14}\, \msol$ , which is essentially
the full galaxy cluster mass range. 

In the following, we focus on a sub-sample of 62 clusters from the ESZ 
catalogue. This sub-sample is defined and extensively described in 
\citet{planck2011-5.2b}, where it was used to calibrate scaling relations 
between the SZ and X-ray cluster properties. All 62 clusters were already known 
in X-rays \cite[found in the MCXC,][]{pif11}, and have good \xmm\ archive data,
allowing for a high quality X-ray data analysis. Masses and radii for the 
sample were estimated using the $\Mv$--$\YX$ relation of A10 \citep[see also 
][]{pra10}, assuming standard evolution (see Eq.~\ref{e:yx}).

The radius $R_{500}$ was calculated iteratively as described in \citet{kra06}. 
While this sample is neither representative nor complete, it is the largest, 
highest-quality SZ--X-ray data set currently available. The majority of objects 
lie at a redshift lower than $0.3$ (and all have $z<0.5$) and cover 
approximately a decade in mass. In angular size, i.e., $\theta_{500}$, they 
range between $3.7$ and $22.8$~arcmin, with a median value of $7.6$~arcmin. 
These extended, SZ-bright clusters are thus ideal targets to investigate the 
spatial distribution of ICM thermal pressure support in clusters of galaxies by 
means of the spatial distribution of their SZ signal.

%
\section{Basic formalism}
\label{s:form} 
%
\subsection{The Sunyaev-Zeldovich effect}
\label{s:sz}

The inverse Compton effect for thermalised electrons under a 
blackbody radiation field is treated in the \cite{kom57} equation. It was 
studied and characterised for the case of thermal electrons in clusters of 
galaxies by \citet{sunyaev70} and \cite{sunyaev72}.
The ensuing spectral distortion of 
the CMB spectrum in the direction of clusters is named the Sunyaev-Zeldovich 
(SZ) effect.

In the following, we leave aside the kinetic SZ effect, a doppler effect, 
resulting from the cluster peculiar motion within the comoving reference frame 
of the Hubble flow. 
Also, the relativistic corrections on the SZ spectrum, of the order of
$kT_{e}/m_{e} c^2$, are not relevant for our study and are thus neglected
\citep[see, e.g.,][]{poi98, challinor98, sazonov98}.

The intensity of the SZ effect is characterised by the dimensionless 
Comptonisation parameter $y$, the product of the average fractional energy
transferred per collision and the average number of collisions:

\begin{equation}
y=\frac{\sigma_{\rm T}}{m_{\rm e} c^2}\int{P(l)dl},
\label{e:yp}
\end{equation}

\noindent where $\sigma_{\rm T}$ is the Thomson cross-section, $m_{\rm e}$
the mass of the electron and $c$ the speed of light.  $P$ is the pressure
produced by the plasma of thermal electrons along the line of sight. 

Assuming the clusters are spherically symmetric, we can express the profile in 
the Comptonisation parameter as a geometrical projection of the spherical
pressure profile along the axis of a cylinder:
\begin{equation}
y(r)  =  \frac{\sigma_{\rm T}}{m_{\rm e} c^2}
 \int_r^{R_{\rm b}}{\frac{2 P(r') r' dr'}{\sqrt{r'^2-r^2}}}.
\label{e:yprof}
\end{equation}

As $y$ is dimensionless we have $y(\theta)\equiv y(r)$. Actual SZ measurements 
derive from the convolution of the $y$-profile on the sky with the 
instrument spatial response, $f_{\rm PSF}$: 

\begin{equation}
\tilde{y}(\theta) = f_{\rm PSF} \otimes y(\theta).
\label{e:psf}
\end{equation}

If the SZ brightness, which is proportional to $y$, is indeed independent of 
redshift, the SZ flux is proportional to the integrated Comptonisation 
parameter and thus depends on the source distance via
\begin{equation}
Y(\theta) = D_{\rm A}(z)^2 Y(R)  =  D_{\rm A}(z)^2
 \frac{\sigma_{\rm T}}{m_{\rm e} c^2}\int_{0}^{R}{2\pi y(r)r dr},
\label{e:ycyl}
\end{equation}
where $D_{\rm A}$ is the angular diameter distance.

In the following, the observed value of the integrated Comptonisation parameter 
is given: in units of Mpc$^2$ when expressed in the source intrinsic reference 
frame, i.e., $Y(R)$; and in arcmin$^2$ when expressed in the \Planck\ satellite 
reference frame, i.e. $Y(\theta)$.  
Analogously we can define the total integrated pressure within a sphere of 
radius $r$, and express it in $Y$ units. We define this 
``pseudo'' or ``spherical'' integrated Comptonisation parameter as
\begin{equation}
Y_{\rm sph}(r) =  \frac{\sigma_{\rm T}}{m_{\rm e} c^2}
 \int_{0}^{R}{4\pi P(r)r^2 dr}.
\label{e:ysph}
\end{equation}

%
\subsection{The scaled pressure profile}
\label{s:ppb}

As mentioned in Sect.~\ref{s:intro}, in a hierarchical scenario of structure 
formation the halo population is self-similar in scale and structure. Profiles 
of physical quantities are universal once scaled according to their
radius and reference quantities defined at a given density contrast, $\delta$.  
Within this self-similar framework, we adopt throughout the paper the value of 
$\delta=500$. 
The scaled pressure profile thus reads

\begin{equation}
\mathbb{P}(x)=\dfrac{P(r)}{P_{\rm 500}} \;\;\;\; , \;\; \textrm{with} 
\;\;\;\;\; x=\dfrac{r}{\Rv}.
\label{e:scale}
\end{equation}

We adopt for the pressure profile the analytical formulation given by 
\citet{nag07} for the generalised Navarro-Frenk White
\citep[GNFW,]{navarro97,nag07} profile

\begin{equation}
\mathbb{P}(x) =  
\dfrac{P_{0}}{(c_{500}x)^{\gamma}
 [1+(c_{500}x)^{\alpha}]^{(\beta-\gamma)/\alpha}},
\label{e:gnfw}
\end{equation}

\noindent where $x=r/\Rv$, and the model is defined by the following 
parameters: $P_0$, normalisation; $c_{500}$, concentration parameter defined at 
the characteristic radius $\Rv$;  and the slopes in the central 
($x\ll1/c_{500}$), intermediate ($x\,{\sim}\,1/c_{500}$) and outer regions 
($x\gg1/c_{500}$), given by $\gamma$, $\alpha$ and $\beta$, respectively. 

A10 have fitted this analytical profile to a combination of observed pressure 
profiles derived for the \rexcess\ sample \citep{boe07} from high quality \xmm\ 
data, together with three sets of predicted profiles from numerical simulations 
of structure formation implementing DM and baryon physics (i.e., radiative 
cooling and recipes for feedback). 
Measurements of the gas mass (integrating the density profile) and the X-ray 
spectroscopic temperature lead to the quantity $Y_{\rm X, 500}= M_{\rm gas, 
500}T_{\rm X, 500}$. This parameter links to the actual integrated parameter as 
seen from the SZ, i.e., $Y_{\rm 500}$, as

\begin{equation}
Y_{500} = A_{\rm XSZ} \dfrac{\sigma_{\rm T}}{m_{\rm m e} c^2}
 \dfrac{1}{\mu_{\rm e} m_{\rm p}}
 Y_{\rm X, 500},
\label{e:yxsz}
\end{equation}
\noindent where  $\mu_{\rm e} = 1.148$, is the mean molecular weight of
electrons for a 0.3 solar abundance plasma and $m_{\rm p}$ is the proton mass.
Here $A_{\rm XSZ}=0.924\pm 
0.004$ from equation~19 of A10 derived from the \rexcess\ sample, and 
$A_{\rm XSZ}=0.95\pm0.04$ from the best fit SZ scaling relation between
$\Yv$ and $Y_{\rm X, 500}$ from \citet{planck2011-5.2b} using the same sample
as in this work. For consistency purposes we use this latest value in the
following. The predicted and measured values for $\Yv$ are compatible within
the $\pm1\sigma$ limits.

$Y_{\rm X, 500}$ is a good proxy for the cluster total mass 
\citep{kra06,nag07,arn07}. The non-standard scaling relation fitted  against 
the \rexcess\ data is provided by their equation~2:
\begin{eqnarray}
&& E(z)^{2/5}\Mv =\nonumber\\
&& \qquad 10^{14.567 \pm 0.010}
 \left[\frac{\YX}{2\times10^{14}\,\msol\,\keV}\right]^{0.561 \pm 0.018}\,
  \msol.
\label{e:yx}
\end{eqnarray}

In turn the characteristic pressure $P_{\rm 500}$ scales with the cluster total 
mass, reflecting the mass variation expected in the standard self-similar 
model, purely based on gravitation (Eq.~5 in A10):
\begin{eqnarray}
&& P_{\rm 500}= 1.65\times 10^{-3} h(z)^{8/3} \nonumber\\
&& \qquad\qquad \times \left[\frac{\Mv}{3\times10^{14} h_{70}^{-1}\, \msol}
 \right]^{2/3}\, h_{70}^{2} \textrm{keV cm$^{-3}$}.
\label{e:pm}
\end{eqnarray}

The $M-Y_{X}$ relation given in Eq.~\ref{e:yx} deviates from the standard
self-similar case (which has a slope of $3/5$). From the definition of $P_{\rm 
500}$, any deviation from the standard self-similar scaling will appear as a 
variation of the scaled pressure profiles. As shown by A10 in their
equation~9, this variation can be expressed as a function of the total mass.
At $\delta=500$, this mass dependence is almost constant with radius, and it
can be approximated by

\begin{equation}
\dfrac{P(r)}{P_{500}} = \mathbb{P}(x) \left[ 
\dfrac{M_{500}}{3\times10^{14}h_{70}^{-1} \msol}\right]^{0.12}.
\label{e:massdep}
\end{equation}

%
\subsection{Conversions and normalisations}
\label{s:conv}
In the following, we measure SZ profiles from the \Planck\ data 
(Sect.~\ref{s:psz}). All through this work, we compare these observed profiles 
with the predictions that arise from the X-ray constraints. For all clusters
we assume the universal pressure profile shape as derived from A10. We 
parametrise it according to the quantities derived from the X-ray analysis  
$\Rv$ and $Y_{\rm X, 500}$ (see previous section). 
We can then derive predicted pressure profiles (see Eq.~\ref{e:gnfw}), and 
Comptonisation parameter profiles (Eqs.~\ref{e:yprof} and~\ref{e:ycyl}). The 
later convolved by the instrument beam (Eq.~\ref{e:psf}) is directly comparable 
to our observed \Planck\ profiles.

We apply a statistical approach to the observed SZ profile and to the pressure 
profile, averaging individual profiles in our sample once scaled. Profiles are 
scaled in radius according to $\Rv$.
The observed SZ profiles (or predicted SZ profiles from X-ray constraints) are 
dimensionless, and we therefore normalise them by the quantity

\begin{equation}
\Phi\equiv \Yv/\Rv^2.
\label{e:phi}
\end{equation}

\noindent In the following we used the value of $Y_{500}$ as defined in 
Eq.~\ref{e:ysph}.
 
To translate our measured SZ profiles, we deconvolve and deproject them for 
each cluster following Eqs.~\ref{e:yprof} and~\ref{e:psf}. In practice this 
operation is performed as described in Sect.~\ref{s:dec}. The resulting 
profiles are converted to pressure such that

\begin{equation}
P(r)=\dfrac{m_{\rm e} c^2}{\sigma_{\rm T}}\dfrac{1}{D_{\rm A}(z)}
 y(\theta)\dfrac{d\theta}{dr}.
\label{e:pnorm}
\end{equation}

%
\section{Reconstruction of the SZ profile}
\label{s:recons} 

\subsection{SZ signal extraction methods}
\label{s:ilc}

The thermal SZ maps were recovered from a combination of \Planck\ channels, 
making use of either real or Fourier space methods.  We applied three methods 
based on internal linear combination (ILC) algorithms to the \Planck\ data: (1) 
the Modified Internal Linear Combination Algorithm  \citep[MILCA, 
][]{hurier10};, (2) Needlet Internal Linear Combination \citep[NILC, 
][]{delabrouille09}; and (3) the Generalized Morphological Component Analysis 
\citep[GMCA, ][]{bob08}. A  description of each method is given in 
App.~\ref{s:met}.

To optimise the reconstructed SZ map we have to take into account that: (i) the 
brightness of the SZ effect (as an increment or decrement) is maximum in the 
sub-millimetre to millimetre range; (ii) the typical angular size of the 
cluster is a few arcmin; and (iii) the final resolution of the reconstructed SZ 
map is determined by the lowest resolution of the combined frequency maps. We 
therefore restricted ourselves to the use of the six \Planck-HFI channels in 
the SZ map reconstruction process. The final resolution of the reconstructed SZ 
maps is that of the lowest-resolution 100\,GHz channel, or $10$ arcmin. SZ maps 
are in units of the (dimensionless) Comptonisation parameter.
Integrated Comptonisation parameters are expressed in arcmin$^2$ for the 
observed values and in Mpc$^2$ for the intrinsic value in the source reference 
frame. The associated SZ noise maps are built in a fully consistent way from 
the individual frequency error maps (see Sect.~\ref{s:data}). 


\subsection{SZ profile computation}
\label{s:psz}

\subsubsection{Profile extraction}

Individual cluster profiles were computed from the reconstructed all-sky SZ 
maps. Uncertainties were obtained from the all-sky reconstructed SZ error maps, 
which were derived by applying the same reconstruction methods to the frequency 
error maps (see Sect.~\ref{s:data}). For instance, in the case of a linear 
method (e.g., ILC based), the coefficients used for the linear combination of 
the frequency maps were propagated to the combination of the frequency error 
maps.

We extracted a square patch of side $20\times \theta_{\rm 500}$ around each 
cluster position from the all-sky SZ map. Patches were projected from {\sc 
HEALPix} \citep{gorski2005} to a tangential projection. For each patch, the 
pixel size was adapted so that it was constant in scaled units of $ \theta_{\rm 
500}$ over the full cluster sample. This unavoidably leads to an oversampling 
of the pixels or, more precisely, to a redundancy of the original all-sky map 
pixels in the reprojected patch. We produced an associated patch that tracks 
this redundancy. We also extracted equivalent patches from the associated 
all-sky SZ error map, and from the all-sky variance map.

We then computed a profile from each SZ map patch. These were calculated on a 
regular radial grid with bins of width $\Delta r/\Rv= 0.25$, allowing us to 
sample each cluster profile with four points within $\Rv$. The $y$ value of a 
bin was defined as the mean of the values of the pixels falling in each 
annulus. We subtracted a background offset from the maps prior to the profile 
computation. The offset value was estimated from the surrounding region of each 
cluster where $r>7\,\Rv$. The uncertainty associated with this baseline 
offset subtraction was propagated into the uncertainty of each bin of the 
radial SZ profile.

\subsection{Expected SZ profile}
\label{s:pred}
We used the GNFW pressure profile shape characterised by A10 from the \rexcess\ 
sample and numerical simulations to generate a pressure model for each cluster 
of our sample. We kept the A10  best fit values of $c_{\rm 500}=1.18$ and 
the three slopes, $\alpha=1.05$, $\beta=5.49$, $\gamma=0.31$. We derived 
the normalisation $P_0$  from the observed value of $Y_{\rm X, 500}$ (see 
Eq.~\ref{e:ysph}), using a conversion of $0.95$ between $Y_{\rm X}$ and $Y_{\rm 
SZ}$ as described in Sect.~\ref{s:ppb}. Finally $\Rv$ was fixed to the values 
derived from the \xmm\ analysis. Both $Y_{\rm X}$ and $\Rv$ are reported in 
\citet{planck2011-5.2b}. We computed the projection matrix and the PSF 
redistribution matrix (for a Gaussian beam of $10$~arcmin FWHM) as expressed in 
 Eqs.~\ref{e:yprof} and~\ref{e:psf}, respectively, to multiply  each pressure 
profile model. The derived $y$ profile model is directly comparable to the 
\Planck\ measurements. The average profile model across the sample was derived 
similarly to the observed stacked profile (see above), and is used to compare 
to the observed stacked profile.


\subsubsection{Accounting for correlation between points}
\label{s:cor}
A certain level of correlation is introduced between the points of a radial 
profile derived in this manner. We account for this correlation in the 
covariance matrix of each profile, which is computed as follows. 

For each map patch, we masked radii $\theta<7\, \theta_{\rm 500}$ centred 
on the cluster and computed the power spectrum of the noise on the remaining 
area. The use of the surrounding regions of the cluster to characterise the 
noise properties allows us to account for the effect of astrophysical 
contamination in the cluster vicinity, as well as the instrument noise and 
systematics. These sources of contamination were, by construction, excluded 
from the all-sky frequency error maps (see Sect.~\ref{s:data}). The power 
spectrum drawn from the SZ error map of each patch is therefore, 
systematically, slightly lower than the one computed over the cluster 
surroundings. 

We then simulated $m=500$ realisations of the noise patch based on this power 
spectrum, accounting for the variance map and sky pixel redundancy and assuming 
an inhomogeneous correlated Gaussian noise. We then extracted a noise profile 
from each realisation, reproducing the baseline background subtraction used for 
the observed profile. The covariance matrix was built from all the simulated 
noise profiles, i.e., $C=P_n^{\rm T}P_n$,
where $P_n$ is an $n\; {\rm points}\times m$ 
matrix of simulated noise profiles. Considering two points as correlated when 
their correlation coefficient is larger than 0.3, the typical level of 
point-to-point correlation in our profiles is about 20\% ($16$\%, 
$21$\% and $28$\% for MILCA, NILC and GMCA, respectively).

\begin{figure*}[!th]
\center
\includegraphics[scale=1.,angle=0,keepaspectratio,width=\textwidth]{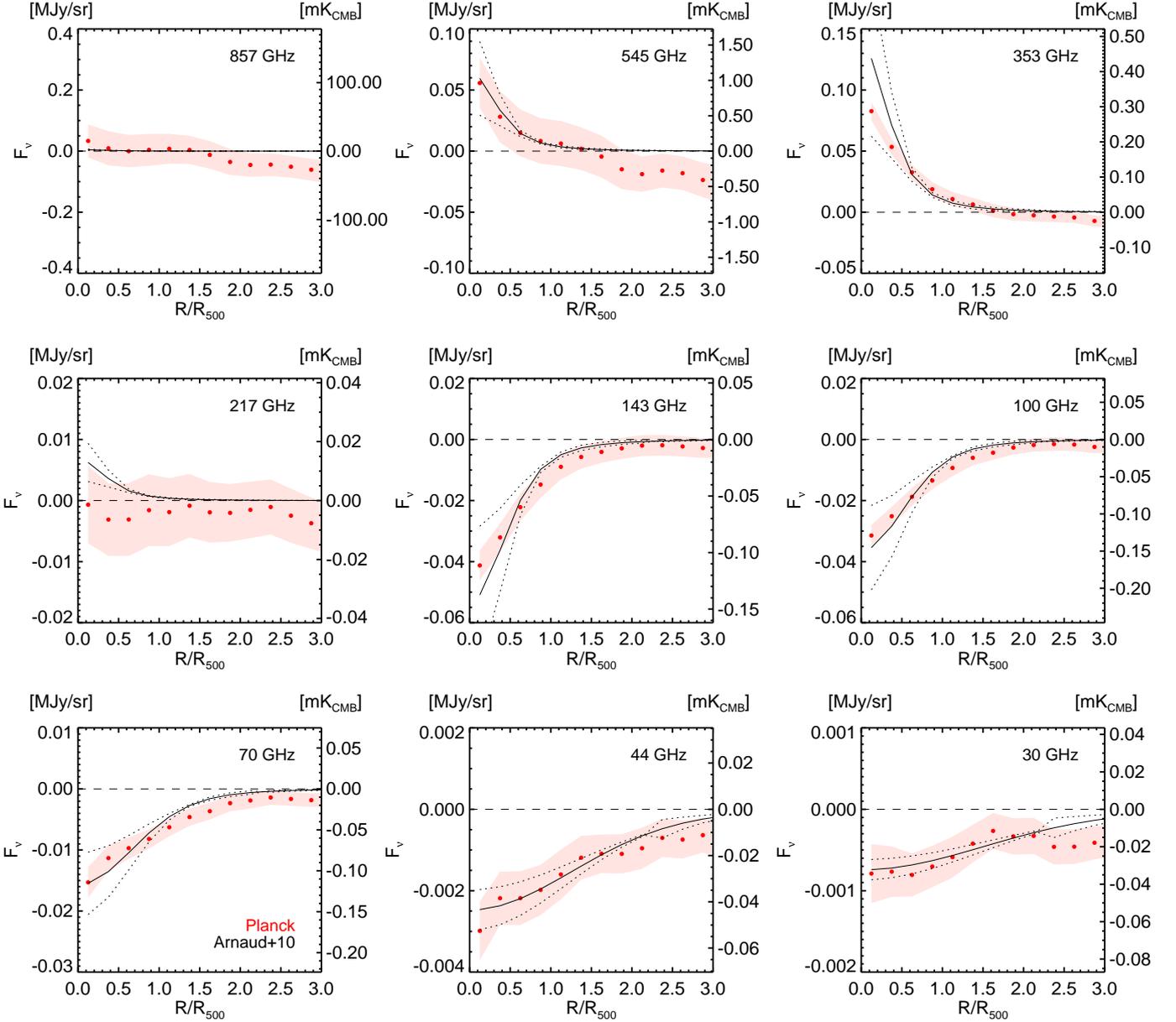}%
\caption{\footnotesize{Average scaled SZ profiles for each of the nine \Planck\ 
frequency bands (decreasing frequency from left to right and from top to 
bottom). The points within each individual profile are correlated at about the 
20\% level (see Sect.~\ref{s:cor}). At each frequency, the stacked 
radial profile is obtained from the average of the 62 individual cluster 
profiles scaled by $\Rv$ and $\Phi_i/\langle\Phi_i\rangle$, in the $x$- and
$y$-axis
directions, respectively (see Sect.~\ref{s:freq}). The light-red shaded area in 
each panel indicates the statistical uncertainty about the average observed 
profile. The solid black line corresponds to the prediction of the universal 
pressure profile (A10) for our cluster sample in each band, and the two dashed 
lines indicate its associated dispersion. }} \label{f:freq}
\end{figure*}


\subsubsection{Stacking procedure}

Given that all profiles are computed on the same grid in scaled radius, they 
can easily be stacked together. Each profile $y_i$ was rescaled by $R_{\rm 500, 
i}$ and $\Phi_i$ (as given in Eq.~\ref{e:phi}), respectively in the X and Y 
axis directions. 
The stacked profile and associated covariance matrix are given by
\begin{equation}
\tilde{y}=\frac{1}{n}\sum_i^n{\frac{y_i}{\Phi_i}}\;\;\;\;\;\; {\rm 
and}\;\;\;\;\;\; \widetilde{C}=\frac{1}{n^2}\sum_i^n{\frac{C_i}{\Phi_i^2}}. \\
\label{e:norm}
\end{equation}

For the computation of $\Phi$, we used $Y_{sph, {\rm 500}}$ in Mpc$^2$ as given 
in Eq.~\ref{e:ysph}. 
For simulated data (App.~\ref{s:sim}), $\Phi$ was derived from the \Planck\ 
$\Mv$--$Y_{{\rm 500}}$ relation \citep{planck2011-5.1b}; for observed data 
(Sect.~\ref{s:pobs}), $\Phi$ was measured directly from the \Planck\ data.

{When for a given bin, values for all clusters are dominated by the signal, we 
assumed a log-normal distribution of their scatter and stacked them in 
logarithmic space. The logarithm of the stacked value and associated error were 
then translated back into linear space. In other terms, for bin $j$, if all the 
clusters satisfied the condition $(y_i^j-\sigma^j_{y,i})>0$, we stacked all the 
measurements in the logarithmic space; otherwise, we did it in the linear 
space}.

We computed the statistical and weighted (i.e., $1/\sigma^2$) average profiles, 
as well as the median profile of the sample, and checked their consistency. 
Over a radial range of $3\times R_{\rm 500}$, the average of the relative error 
of the weighted mean and median profiles with respect to the statistical 
average profile is smaller than $10$\% and $5$\%, respectively. Based on this 
agreement, we used the statistical average to compute the stacked profile 
throughout this study.

Of the 62 clusters in our ESZ--{\it XMM\/} sample, only two are 
spatially close and thus are potentially physically connected: A3528 and A3532. 
These two clusters are members of the Shapley supercluster. Each of those
clusters was masked over an area of radius $3\,\Rv$ when processing the other.
Thus, we consider that their profiles and covariance matrices are independent. 

At the end of our data processing, the stacked covariance matrix encompasses 
the statistical errors due to instrumental noise, astrophysical fluctuations at 
the cluster location and systematic effects (e.g., instrumental and arising
from data 
processing).  For the purposes of our study we have also computed the 
dispersion across our sample for each position within our profiles.
Both uncertainties are propagated throughout the analysis, and are
cross-compared in the following sections. 
{In Figures \ref{f:met}, \ref{f:ppp} and \ref{f:psim} the error bars shown on the \Planck\ data points are purely statistical and correspond to the square root of the diagonal elements of the covariance matrix (i.e., Eq~\ref{e:norm}). The rms scatter of individual profiles around the mean is indicated by a coloured band.}

%
\section{\textit{Planck\/} galaxy cluster SZ profile}
\label{s:szprof}

%
\subsection{Frequency stacked profiles}
\label{s:freq}

We first looked for the raw cluster signature in each of the nine \Planck\  
frequencies. We extracted a brightness profile at each frequency from the raw 
\Planck\ maps for each of the 62 clusters in the ESZ--{\it XMM\/} sample. These 
were rescaled in units of $\Rv$ in the $x$-axis direction and in units of 
$\Phi_i/\langle\Phi_i\rangle$ in the $y$-axis direction
(see Sect.~\ref{s:conv}), 
and then stacked. The resulting average brightness profiles are shown in 
Fig.~\ref{f:freq}, where the shaded area in each panel depicts the statistical 
error about the mean flux value in each bin.

The cluster signal appears clearly in most of the nine frequencies, following 
the SZ thermal spectral signature. The signal is positive at 545 and 353\,GHz, 
compatible with zero at 217\,GHz, and then negative down to 30\,GHz. 
The 62 clusters of our sample are strong SZ sources, as expected for 
objects detected at $S/N>6$ \citep[][]{planck2011-5.1a}. The stacking procedure 
mostly averages out the effect of foreground and background contaminants; 
however, they contribute to the final dispersion in the stacked profiles. Owing 
to their Gaussian nature, CMB fluctuations are more easily washed out in the 
stacking procedure than the Galactic and extragalactic dust emission in the 
high frequency range, and Galactic and point source radio emission (i.e., 
free-free and synchrotron) in the lower frequency  range. 

We also plotted for each frequency the expected profile computed as described 
in Sect.~\ref{s:pred}. This comparison shows a first order agreement between 
the \Planck\ measurements and the expected SZ profiles derived from X-ray 
constraints assuming a GNFW pressure profile shape.

%
\subsection{Stacked SZ image}
\label{s:imsz}

\begin{figure}[!t]
\center
\includegraphics[scale=1.,angle=0,keepaspectratio,width=\columnwidth]{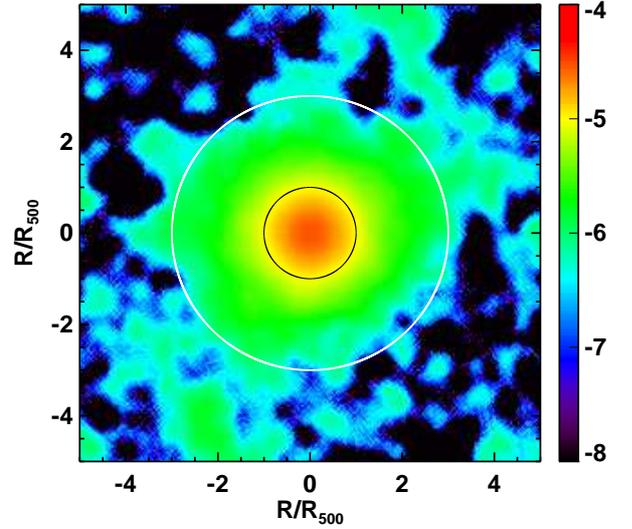}%
\caption{\footnotesize{
Stacked scaled image (size is $10\,\Rv$ on a side) using a 
logarithmic stretch in $y$. Individual maps are rescaled by $\Phi_i$ before
averaging, and then multiplied by $\langle\Phi_i\rangle$.  The black and white
circles mark the loci of $1$ and $3\times \Rv$, respectively.}}
\label{f:im}
\end{figure}

Figure~\ref{f:im} presents the stacked average image of the 62 clusters 
of our ESZ--{\it XMM\/} sample, plotted on a logarithmic scale. Before 
averaging, each individual SZ map was normalised identically to the profiles 
(see Eq.~\ref{e:phi}) and randomly rotated by $0^\circ$, $90^\circ$,
$180^\circ$ or $270^\circ$. The stacked image was then renormalised by
$\langle\Phi_i\rangle$ in order to be expressed in $y$ units.
The image shows that the SZ signal is detected out to ${\sim}\, 3\times \Rv$, 
marked by the white circle in the figure. The extent of existing observational 
constraints on the gas distribution in clusters in SZ and X-rays, i.e., $\Rv$, 
is marked by the black circle.  We also generated a jackknife-type map by 
averaging the scaled individual SZ maps, $m_i$, as $\sum{(-1)^i m_i}$. The rms 
values of both the stacked SZ and  {\it on-off} maps beyond $5\times \Rv$ are 
compatible: $4.6\times 10^{-7}$ and $4.8\times 10^{-7}$, and, in turn, 
compatible with the rms value below this radius on the jackknife stacked map:
$3.5\times 10^{-7}$. This check demonstrates that our average map is not
strongly affected by residuals or biases. The radial profile computed from
the stacked map is fully compatible with the stacked profile discussed below.

%
\subsection{Observed stacked SZ profile}
\label{s:pobs}

\begin{figure*}[!th]
\center
\includegraphics[scale=1.,angle=0,keepaspectratio,width=0.95\columnwidth]
 {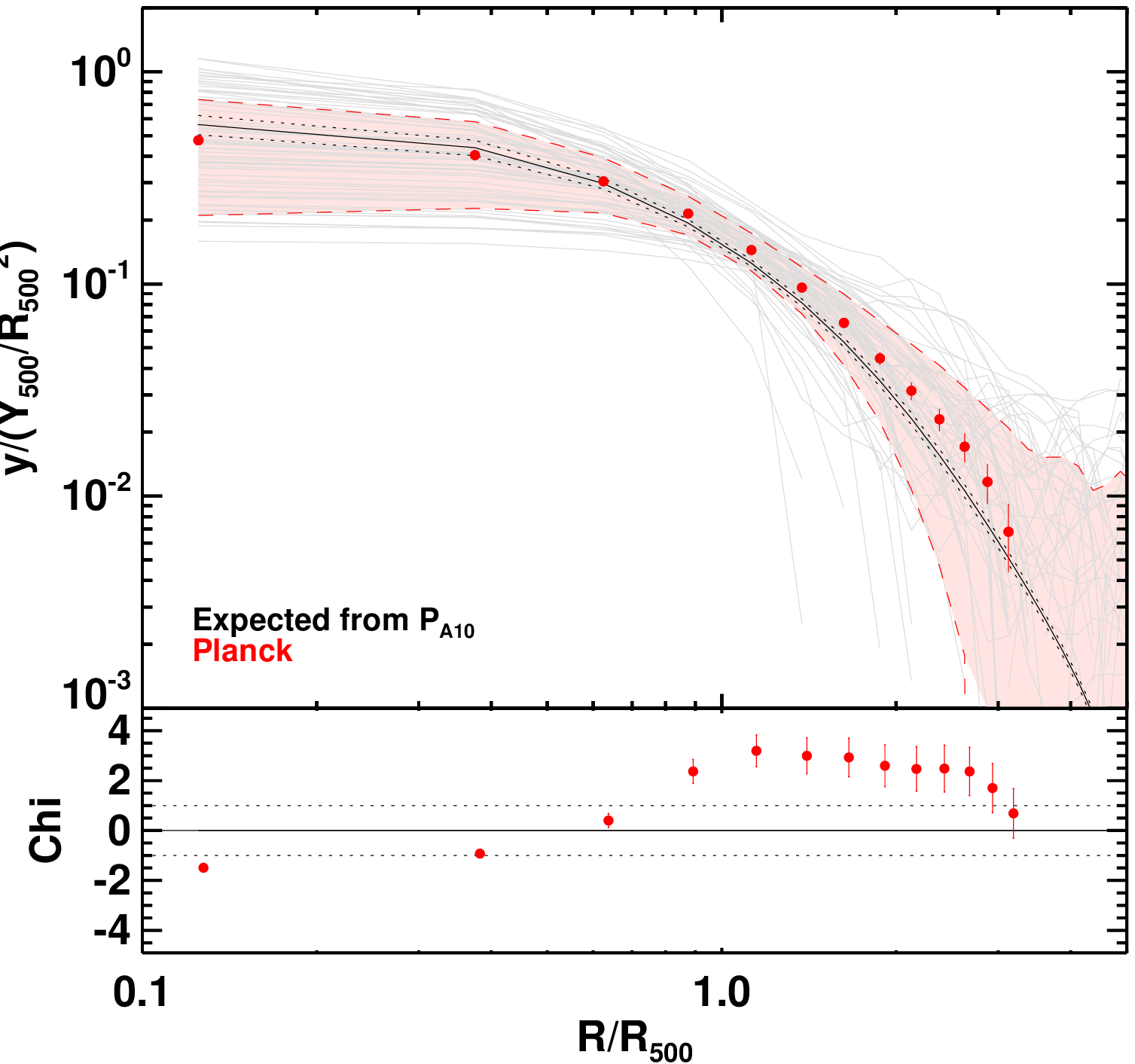}%
\includegraphics[scale=1.,angle=0,keepaspectratio,width=1.06\columnwidth]
 {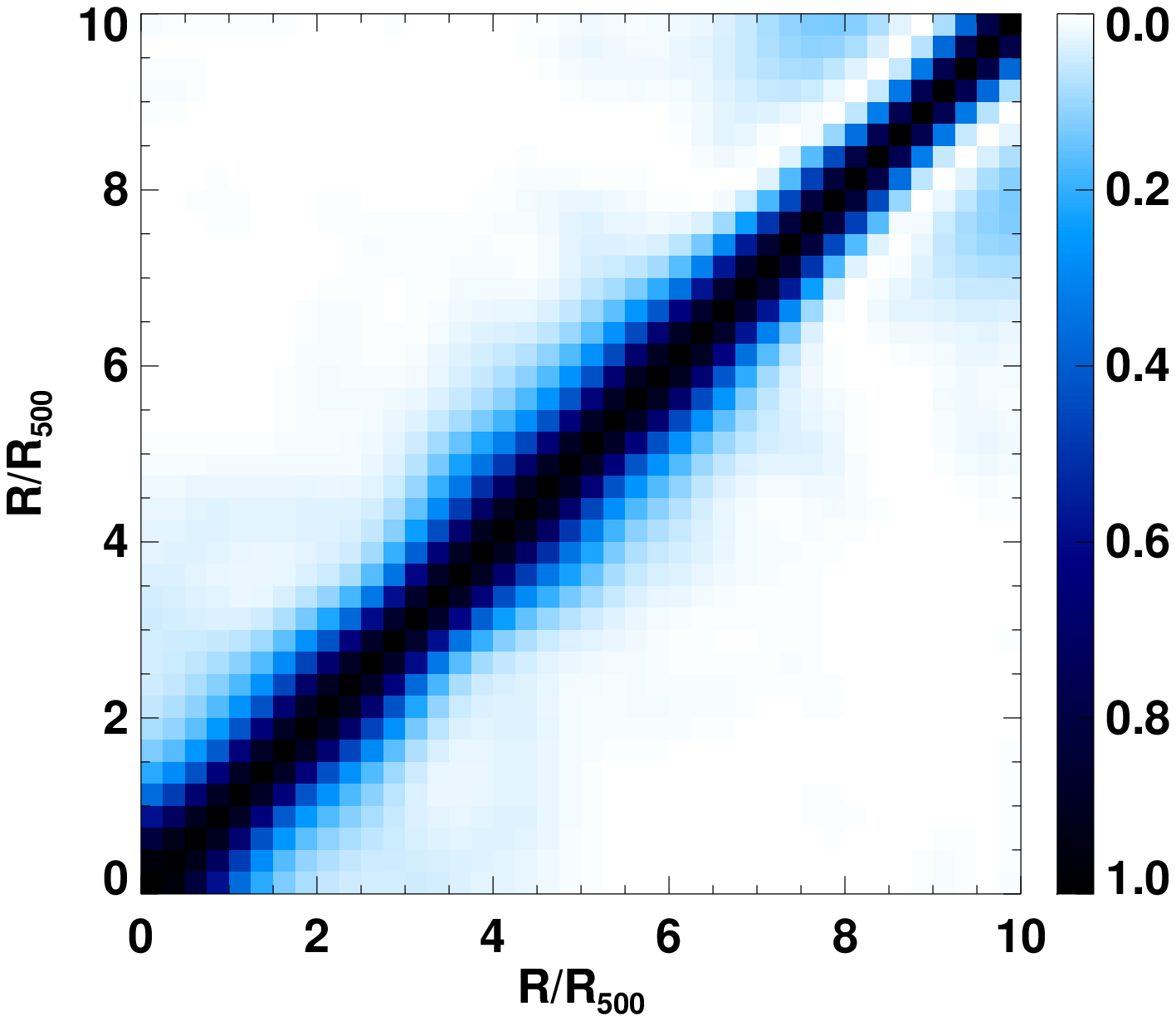}
\caption{\footnotesize{
Left: individual profiles (grey lines) and scaled stacked profile (red 
points) for our sample of 62 clusters.  The light-red shaded area marks the 
dispersion about the average stacked profile, with its upper and lower limits
highlighted by the two dashed red lines. The points within each individual 
profile are correlated at about the 20\% level (see Sect.~\ref{s:cor}).  {The  error bars on the \Planck\ data points are purely statistical and correspond to the square root of the diagonal elements
of the covariance matrix.}
The solid black line (labeled in the legend as ``Expected from $P_{A10}$'') is 
the average stacked profile obtained from the expected SZ individual profiles 
drawn from the universal pressure profile of A10, parameterised according $\Rv$ 
and $\Yv$ derived from the \xmm\ data analysis presented in  
\citet{planck2011-5.2b}. The two dotted black lines indicate the associated 
dispersion about this model profile. The bottom panel shows the value of $\chi$ 
at each point with respect to the expectations  from the universal profile 
taking into account the measured errors. Right: matrix of the 
correlation coefficient for the observed radial stacked profiles. Both panels 
are illustrated here for the MILCA results.}}
\label{f:met}
\end{figure*}

We derived the observed stacked \Planck\ SZ profiles for the ESZ--{\it XMM\/}
sample as described in Sect.~\ref{s:psz}.
$\Yv$ is obtained from the algorithms used for blind detection of SZ clusters 
in the \Planck\ survey \citep{planck2011-5.1a}, namely the Powell Snakes  
\citep[PwS,][]{car09b,carvalho11} and multi-frequency matched filter 
\citep[MMF3,][]{mel06} algorithms. In both cases the algorithms implement a 
universal pressure profile shape (A10) with the position of the cluster fixed 
to the \xmm\ coordinates and the size of the universal pressure profile fixed 
to $\Rv$. The fluxes from both methods (i.e., PwS and MMF3) are consistent over 
the whole sample. The median value for the ratio PwS/MMF3 is $0.96\pm 0.05$ 
(see App.~\ref{s:ymes}). To further validate the above, and as a consistency 
check with previous \Planck\ results, we fitted each individual SZ profile with 
a projected, PSF-convolved universal pressure profile. We fixed $\Rv$ to the 
best fitting X-ray value from \citet{planck2011-5.2b} and only fitted the 
normalisation, $Y_{\rm 500}$. 
 Given the agreement, we used the MMF3 values to compute $\Phi$  for each 
cluster (Eq.~\ref{e:phi}).

Appendix~\ref{s:sim} presents a detailed investigation of the convergence 
between stacked profiles derived from each of the three different methods of SZ 
signal reconstruction. We find that all three methods agree remarkably 
well over the entire radial range, both for simulated and observed SZ profiles. 
The three SZ reconstruction methods lead to profiles fully compatible with each 
other. Across the range of radii over which the profiles have been computed 
(i.e., [0--10]$\times\Rv$), accounting for the correlated errors of each 
profile, the reduced $\chi^2$ of NILC and GMCA with respect to MILCA are 
$0.48$ and $0.62$,
respectively\footnote{As our three reconstruction methods are 
remarkably consistent, for the clarity of display and discussion, hereafter we 
illustrate our presentation with one of the three methods only. In each case
the specific method used will be indicated.}. 

The left panel of Fig.~\ref{f:met} shows the stacked MILCA profile for the 62 
clusters in the ESZ--{\it XMM\/} sample compared to the individual profiles 
(see Sect.~\ref{s:psz}). 
The SZ signal is statistically detected over more than two decades in 
intensity, and out to a remarkably large radius of ${\sim}\,3\,\Rv$, reaching 
far into the cluster outskirts. Assuming that the virial theorem can be applied 
(which at such large radius might be breaking down), the outer radius of our 
 statistical detection corresponds to a density contrast of
$\delta\,{\sim}\,50$. 
More conservatively, we can assume that we are statistically probing the 
average SZ and pressure distribution down to regions of density contrast of 
$\delta\,{\sim}\,50$--100. The dispersion about the mean profile dominates the 
statistical uncertainties. It is minimal (by construction) at ${\sim}\,20\%$
around $\Rv$, but increases towards the centre and the outskirts to
${\sim}\,50$\% and ${\sim}\,65$\%
at $0.3$ and $2\times \Rv$, respectively. At the most external radius of our
statistical detection, i.e., ${\sim}\,3\times \Rv$, the dispersion of 
the individual profiles about the mean is more than $100\%$, as at these large 
radii, the individual SZ profiles are fully dominated by noise.

The right panel of Fig.~\ref{f:met} shows the correlation coefficient matrix 
for the stacked profiles presented on the left panel, and illustrates the 
degree of correlation between the points in the profiles.

%
\subsection{Comparison with expectations from universal profile}
\label{s:xsz}

The stacked model used for comparison was computed as described in 
Sect.~\ref{s:pred}.
When considering the statistical errors only, the measured stacked profile is 
significantly above the model.
Taking into account the error on the model and the correlated errors between 
points of the reconstructed SZ profile, we obtained a reduced $\chi^2$ value 
of $3.53$
within $3\,\Rv$. If we omit the error on the model, this value becomes 7.48
As our tests on simulations show a very good agreement between the input model 
and the output reconstructed profiles (see App.~\ref{s:sim}), the difference 
observed here between the measured SZ profiles and the predicted model from 
X-ray constraints is not an artefact of the method, but an intrinsic 
difference. This difference is significant at a 2--3$\sigma$ level from around 
$\Rv$ out to $3\,\Rv$.

%
\section{The galaxy cluster pressure profile from \textit{Planck}}
\label{s:pprof}

\begin{figure*}[!t]
\center
\includegraphics[scale=1.,angle=0,keepaspectratio,width=\columnwidth]{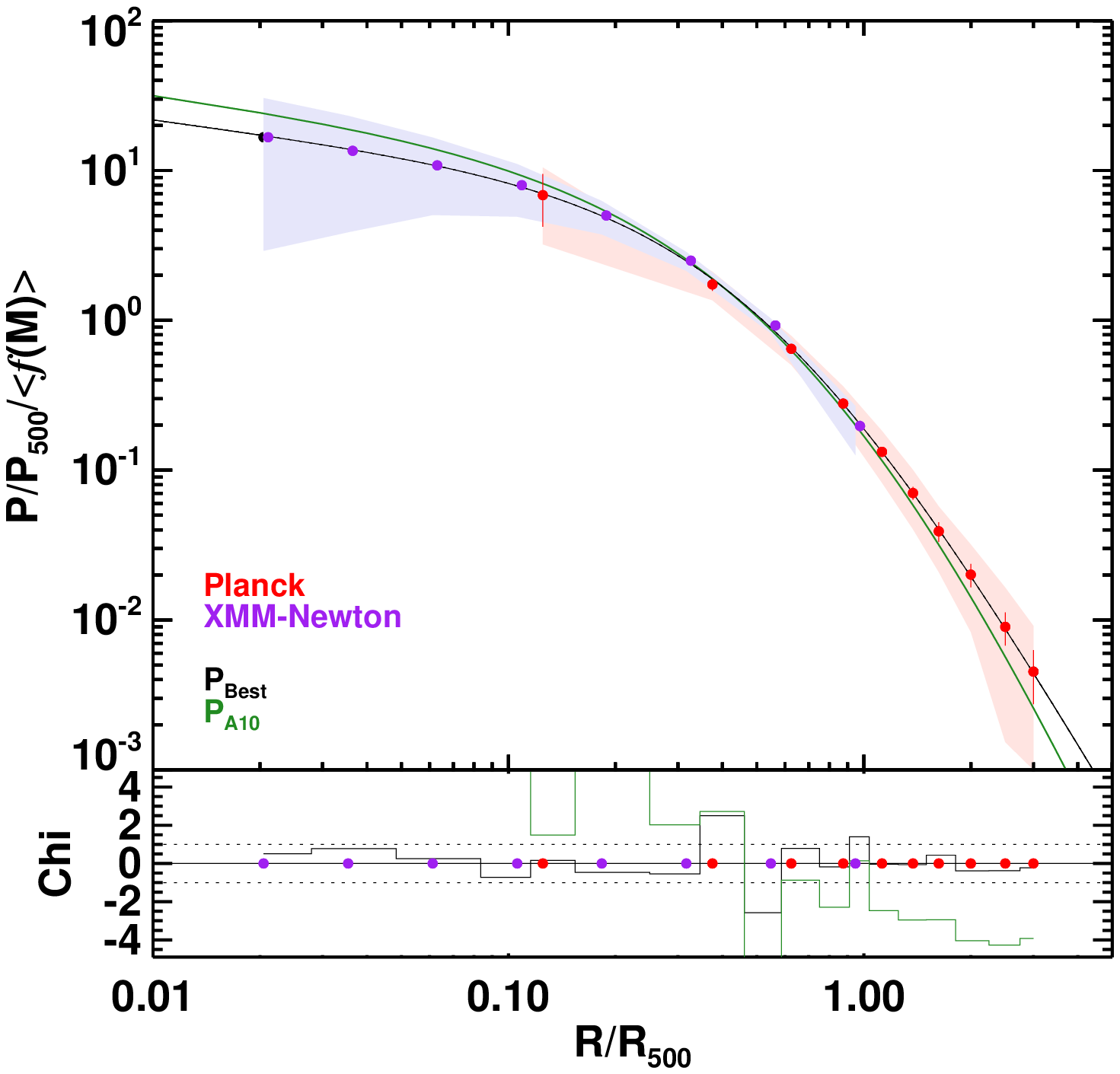}
\includegraphics[scale=1.,angle=0,keepaspectratio,width=\columnwidth]{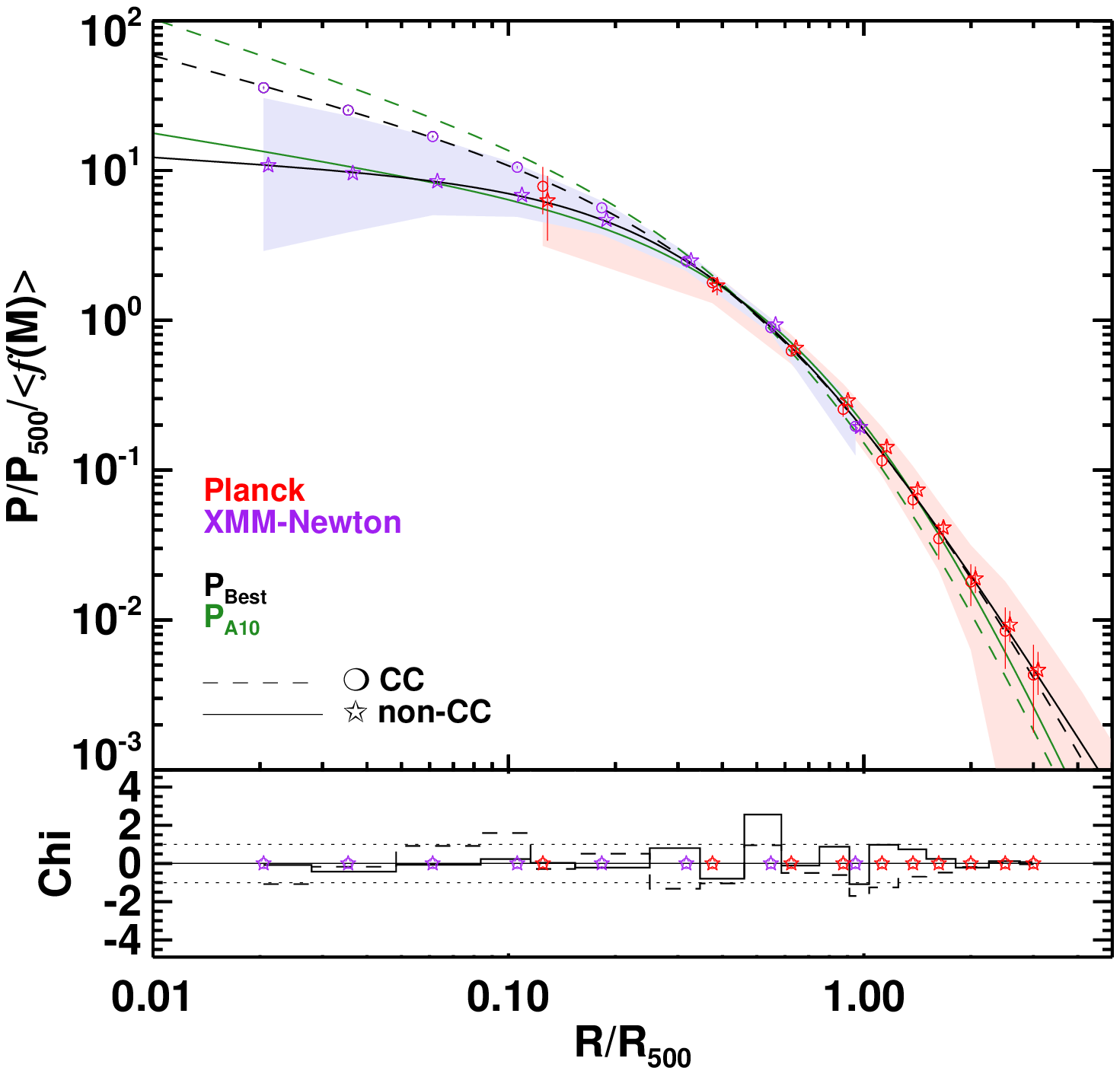}
\caption{\footnotesize{Left: \Planck\ pressure profile obtained from the 
average of the individual pressure profiles across our sample (red points) 
shown together with the stacked pressure profile derived from the \xmm\ data 
for the same sample (purple points). The dispersions about the SZ and X-ray 
profiles are depicted, respectively, by the red and purple shaded areas. Our 
best fit GNFW profile is shown as a solid black line and that of A10 as a solid 
green line. The lower panel shows the $\chi$ profile of these two best models 
taking into account the statistical errors and dispersion about the observed 
profile. Right: stacked profile from \Planck\ and \xmm\ for the 
sub-samples of cool-core (open circles) and non cool-core (open stars) clusters 
within the ESZ--{\it XMM\/} sample. The shaded areas are identical to the one 
shown in the left panel. Our best fit models for each sub-sample are shown as 
black solid and dashed  lines for the cool-core and non cool-core clusters, 
respectively (see Table~\ref{t:fit}). The best A10 fit for cool-core and non 
cool-core clusters are shown as green solid and dashed lines, respectively. The 
lower panel gives the $\chi$ profiles of our best fit models with respect to 
the measured profiles and associated errors (including dispersion). On both 
panels the \Planck\ data points are correlated at about the 20\% level (see 
Sect.~\ref{s:cor}). 
{As for the stacked SZ profile, the error bars on the \Planck\ points are purely statistical and correspond to the square root of the diagonal elements of the covariance matrix. For the \xmm\ points they correspond to the statistical error on mean.}
}}
\label{f:ppp}
\end{figure*}

%
\subsection{PSF deconvolution and deprojection}
\label{s:dec}

The pressure profile is derived for each cluster in our sample by applying a 
deconvolution and deprojection to the observed SZ profiles. As already
mentioned, we assume spherical symmetry. The convolution by the instrumental
beam and the geometrical projection of a spherical pressure profile into a
Comptonisation parameter profile are expressed in Eqs.~\ref{e:psf} and
\ref{e:yprof}, respectively. 
We have applied a deconvolution plus deprojection algorithm adapted from the 
method described by \citet{croston06}. This method allows us to perform a 
straightforward real space deconvolution and deprojection using a 
regularisation procedure originally applied to X-ray surface brightness 
profiles to derive the underlying density profile. The method is adapted to 
lower statistics (i.e., lower number of bins per radial profile).
In the present case, the PSF response matrix was computed for a circular 
Gaussian beam with ${\rm FWHM}=10$~arcmin,
i.e., the angular resolution of our SZ 
maps. SZ cylindrical profiles deconvolved from the PSF  were then deprojected 
into 3D  profiles (assuming spherical symmetry for the clusters), and converted 
to pressure following Eq.~\ref{e:pnorm}.

We used a Monte Carlo (MC) method to propagate the measured error over the SZ 
profiles onto the PSF-corrected and deprojected profiles, accounting for the 
covariance between points expressed in the form of the matrix $C_i$ for the 
cluster $i$. For each cluster $i$, its covariance matrix was Cholesky 
decomposed (i.e., $C_i=L_i L_i^{\rm T}$).
We assumed correlated Gaussian noise, and generated $m=10{,}000$
realisations of the SZ profile, $P_i$, with respect to this 
decomposition (i.e., $\tilde{P}_i^j=P_i+x\, L_i$, where $x$ is an array of 
random numbers following a normal distribution). With this method we make sure 
we properly sample the noise properties of each profile. Each realisation was
then rebinned logarithmically in radius and run through our deconvolution and 
deprojection algorithm. We conservatively chose to feed to the 
algorithm values of the weight per point equal to the inverse of the square of 
the flux dispersion in this bin (i.e., $w=1/\sigma_{\rm flux}^2$). The weights 
remained the same for all realisations. These $m$ realisations of the pressure 
profile were used to compute the covariance matrix of the cluster pressure 
profile similarly to what is described in Sect.~\ref{s:cor}. With this MC 
process, we ensure a proper propagation of the correlated noise through the 
PSF-correction and deprojection of the SZ profile.

%
\subsection{The stacked pressure profile}
\label{s:ppp}
As we did for the SZ profiles, the 62 pressure profiles were rescaled in
the $x$- and $y$-axis
directions according to $\Rv$ and $P_{\rm 500}$, and then stacked together. 

To further compare with A10's results, we accounted for the slight difference in mass range between the \rexcess\ sample and ours via the factor $f(M)=(M_{500}/3\times10^{14}\,h_{70}^{-1}\, \msol)^{0.12}$  \citep[Eq.~\ref{e:massdep}, and see also ][]{sun11}.  We divided our stacked profile by the average value across our ESZ--{\it XMM\/} sample, i.e., $\langle 
f(M)\rangle=1.09$. 

Our three reconstruction methods lead to compatible stacked pressure profiles. 
However, we have accounted for their fluctuations by adding to the diagonal 
terms of the covariance matrix of the stacked profile the maximum 
point-to-point difference between the MILCA profile and the two other methods.
In parallel to our \Planck\ SZ analysis, we have also derived the pressure 
profiles of our 62 clusters from the analysis of  \xmm\ data. We followed the 
method applied to the \rexcess\ sample and presented in A10. Details of the 
\xmm\ analysis and results are provided in \citet{dem12}.

The \Planck\ and \xmm\ stacked pressure profiles derived from our sample of 
62 \xmmesz\ clusters, are displayed in Fig.~\ref{f:ppp}. The two sets of 
data agree remarkably well. They are fully compatible within their respective 
dispersions, and they overlap in the radial range ${\sim}\, (0.3{-}1)\,\Rv$. We 
recall that the points from the \Planck\ data suffer a degree of correlation at
about the 20\% level,
whereas all the points in the \xmm\ profile can be considered as uncorrelated.
These two independently derived pressure profiles are fully complementary. 
Indeed, in the central region the pressure profiles are very well constrained 
from the X-ray data, whilst the \Planck\ measurements hardly reach down to
$0.1\, \Rv$ due to the moderate spatial resolution of our SZ maps.
At larger radii, the extent of the X-ray observations is limited to radii 
smaller than ${\sim}\, (0.7{-}1.0)\,\Rv$ because of the \xmm\ field of view and
of the quick drop in X-ray surface brightness with increasing radius.
The \Planck\ profile extends far beyond this radius. 

With this joint constraint, we bring for the first time a comprehensive 
observational view of the distribution of the average thermal pressure 
distribution in clusters of galaxies  out to a density contrast of
$\delta{\sim}\,50$--100. 

Following A10, we have investigated two sub-samples of our ESZ--{\it XMM\/} 
clusters. Keeping the cool-core (CC) versus non cool-core (non-CC) 
classification as provided in \citet{planck2011-5.2b}, we computed the stacked 
profiles for the  22 CC versus 40 non-CC clusters. Both stacked profiles are 
displayed in the right panel of Fig.~\ref{f:ppp}. They are, as expected, 
different in the central parts, with a more peaked profile for CC systems and a 
shallower one for non-CC clusters. However, in the outer parts, our observed 
profiles for the two subsamples have very similar slopes.

\begin{figure*}[!t]
\center
\includegraphics[scale=1.,angle=0,keepaspectratio,width=0.98\textwidth]{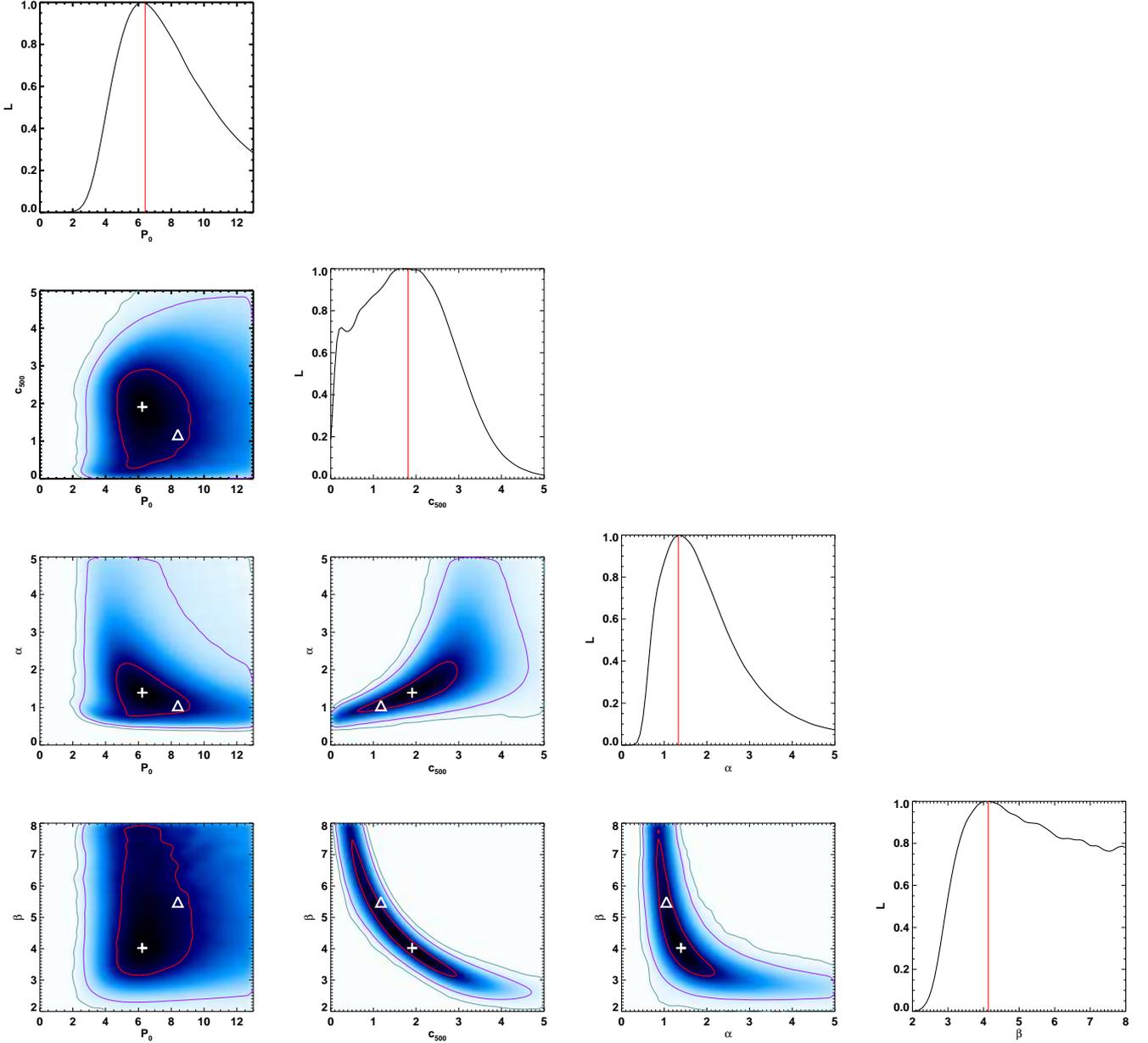}
\caption{\footnotesize{Marginalised posterior likelihood  distribution for our 
best MCMC fit to a GNFW pressure profile with free parameters [$P_0, c_{\rm 
500}, \alpha, \beta$] and $\gamma=0.31$. The best fit values of the parameters
are marked as white crosses and solid red lines, respectively, on the 2D and 1D 
posterior distributions. On the 2D distributions, the white triangles mark the 
value of the A10 parameters, while the red, purple and green solid lines define,
respectively, the 68, 95 and 99\% confidence levels.}}
\label{f:like}
\end{figure*}
%
\subsection{Best-fit to the Generalised NFW profile}
\label{s:fit}
We have combined the \Planck\ and \xmm\ pressure profiles into a joint fit to 
the generalised NFW model (Eq.~\ref{e:gnfw}).
The pressure profiles derived over our sample from the \Planck\ and \xmm\ data 
cover a radial range from $0.02\times \Rv$ to $3\times\Rv$, with an overlap in
the radial interval of $(0.1{-}1)\times\Rv$. 
In order to find the best parametric representation of our observed stacked 
profile, we ran a Monte Carlo Markov Chain analysis to find the maximum 
likelihood solution parametrising the GNFW profile. The \Planck\ and \xmm\ 
data are two different and fully independent observational data sets. We 
therefore computed the likelihood as the product of the two independent 
likelihoods: $\mathcal{L}=\mathcal{L}_{\rm X}\times\mathcal{L}_{\rm SZ}$.
For the \Planck\ data, we have accounted for the correlation 
between points through the use of the covariance matrix (see 
Sect.~\ref{s:dec}). We also accounted for the dispersion about the average 
X-ray and SZ  profiles across the sample by quadratically adding it to the 
errors bars of the \xmm\ points and to the diagonal elements of the covariance 
matrix for the \Planck\ profile. 

We performed our MCMC analysis in log--log space for various combinations 
of free parameters. Fixed parameters were assigned the values from the best fit 
provided by A10. Our best fit parameters are reported in Table~\ref{t:fit}. 
In order to find the best possible analytical representation of our data, we 
assessed the quality of the fit via the value of the reduced $\chi^2$ using 
only the statistical errors, $\bar{\chi}^2_{\rm stat}$. The configuration with
four and five free parameters are of equivalent quality, the former being
slightly better (i.e., $\bar{\chi}^2_{\rm stat}=0.9$ and $1.0$, respectively).
Unsurprisingly the degeneracy between free parameters increases with their
number, translating into a drastic increase in the parameter covariance.
We therefore adopted the four free parameter approach as our best
parametrisation of a GNFW profile of our data, 
i.e., [$P_0, c_{\rm 500}, \alpha, \beta$] free, with $\gamma$ fixed to $0.31$. 
The associated values are marked in bold face in Table~\ref{t:fit}{, and are 
given purposefully with high precision with respect to the errors on 
parameters. These uncertainties are illustrated by the corresponding
marginalised 
posterior likelihood distributions that} are shown for individual (1D) and 
pairs of parameters (2D) in Fig.~\ref{f:like}.

\begin{table}[!tmb]
\begingroup
\newdimen\tblskip \tblskip=5pt
\caption{Best fit parameters for a generalised NFW pressure profile, with
different numbers of fixed parameters.  The $\bar{\chi}^2_{\rm stat}$ value
is the reduced $\chi^2$
computed with respect to the statistical error bars (i.e., not including the
dispersion across the sample). The bold-face line indicates the best parametric
representation of our combined \xmm\ and \Planck\ pressure profile (see
Sect.~\ref{s:fit} for further details). The number of degrees of freedom is
given in the last column.}
\label{t:fit}
\nointerlineskip
\vskip -3mm
\footnotesize
\setbox\tablebox=\vbox{
  \newdimen\digitwidth 
  \setbox0=\hbox{\rm 0} 
   \digitwidth=\wd0 
   \catcode`*=\active 
   \def*{\kern\digitwidth}
   \newdimen\dagwidth 
   \setbox0=\hbox{$^\dagger$} 
   \dagwidth=\wd0 
   \catcode`!=\active 
   \def!{\kern\dagwidth}
\halign{\hfil#\hfil\tabskip=2.4em&
\hfil#\hfil&
\hfil#\hfil&
\hfil#\hfil&
\hfil#\hfil&
\hfil#\hfil&
\hfil#\hfil\tabskip 0pt\cr
\noalign{\doubleline\vskip -2pt}
$P_0$ & $c_{500}$ & $\gamma$  & $\alpha$ & $\beta$ & $\bar{\chi}^2_{\rm stat}$
 & $N_{\rm dof}$\cr
\noalign{\vskip 2pt\hrule\vskip 2pt}
\multispan7 All clusters\hfil\cr
\noalign{\vskip 2pt\hrule\vskip 2pt}
$*6.32$& $1.02$& $0.31^\dagger$& $1.05^\dagger$& $5.49^\dagger$&$   
 3.8$& 15\cr
$*6.82$& $1.13$& $0.31^\dagger$& $1.05^\dagger$& $5.17$!&
 $5.8$& 14\cr
$\bf *6.41$& $\bf 1.81$& $\bf 0.31^\dagger$& $\bf 1.33$!&
 $\bf 4.13$!& $\bf 0.9$& \bf 13\cr
$*5.78$& $1.84$& $0.35$!& $1.39$!& $4.05$!& $1.0$& 13\cr
\noalign{\vskip 2pt\hrule\vskip 2pt}
\multispan7 Cool-core clusters\hfil\cr
\noalign{\vskip 2pt\hrule\vskip 2pt}
${11.82}$& $ 0.60$& $0.31^\dagger$& $0.76$!& $6.58$!& $1.1$&\cr
\noalign{\vskip 2pt\hrule\vskip 2pt}
\multispan7 Non cool-core clusters\hfil\cr
\noalign{\vskip 2pt\hrule\vskip 2pt}
$*{4.72}$& $2.19$& $0.31^\dagger$& $1.82$!& $3.62$!& $1.2$&\cr
\noalign{\vskip 2pt\hrule\vskip 2pt}
}}
\endPlancktable
\vskip 1mm
\tablenote {\dagger} {Fixed parameters which are assigned the best fit 
values of the A10 profile.}\par
\endgroup
\end{table}

In parallel, following exactly the same procedure, we also fitted the stacked 
profiles for the sub-samples of cool-core and non cool-core clusters. The best 
fit parameters are also reported in Table~\ref{t:fit}. Both the stacked 
profiles and best fit model are shown on the right panel of Fig.~\ref{f:ppp}.

We further adopted this four free parameter configuration to fit each of our 
measured individual pressure profiles, combining  \xmm\ and \Planck\ pressure 
data. As describe in the above section for the stacked profile, we accounted 
for the  point-to-point differences between our three reconstruction methods 
(see Appendix~\ref{s:met}) in the error budget. Results are reported in 
App.~\ref{s:ind}, Fig.~\ref{f:ind} and Table~\ref{t:ind}.

%
\section{Discussion}
\label{s:dis}

%
\subsection{The observed pressure profile}
\label{s:compobs}
\subsubsection{The core of clusters}
Our present results are in very good agreement with the universal profile 
derived by A10 outside the core and down to $\Rv$. Within the core, i.e., 
$R<0.15\,\Rv$, our observed profile lies significantly below the A10 
profile.  
In comparison to the \rexcess\ sample which is X-ray selected, our
ESZ--{\it XMM\/} sample is SZ selected (except for the
intersection with the \xmm\ 
archives) and is thus, a priori, closer to being a mass-selected 
sample. It therefore contains more dynamically disturbed clusters (e.g., A2163, 
RXJ2228+2036, etc.). As has also been indicated by results in the validation 
follow-up of \Planck\ clusters with \xmm\ 
\citep{planck2011-5.1b,planck2012-I,planck2012-IV}, this seems to indicate that 
X-ray selection under-samples the population of morphologically disturbed 
clusters. 
When splitting our sample into CC and non-CC clusters, the average profile for 
the CC clusters (and its associated best fit) is in better agreement with the 
universal profile given in A10. However, it is still slightly lower in the
central parts than the A10 profile for CC clusters (see right panel of 
Fig.~\ref{f:ppp}). There might thus still be differences between the population 
sampled by our CC clusters and those from A10. On the contrary, our non-CC 
cluster profile agrees well with the A10 non-CC profile, except for the very 
central parts ($R<0.04\,\Rv$).

\subsubsection{The inner profile}
Observationally, it is hard to accurately determine the
distribution of the SZ signal and, thus, of the 
underlying thermal pressure. Early single target SZ studies were limited to
the clusters' inner regions (i.e., within $\Rv$), and were not really 
competitive with the X-ray measurements in terms of resolution and sensitivity 
\citep[see e.g.,][]{poi01, halverson09, korngut10}. However, the combination of 
SZ and X-ray tracers has already shown its potential in terms of structural 
studies of clusters \citep{poi02,kit04, jia08,basu10}. The first studies 
working on SZ cluster samples were presented by the SPT and ACT collaborations 
\citep{pla10,seh10}, with 15 and 9 high significance clusters, respectively.
The SPT team found good agreement between the X-ray predicted signal and their
SZ measurement within $\Rv$. Similarly, in a recent study, \citet{bonamente11} 
used  a sample of 25 clusters observed with SZA/CARMA \citep{muc07} and  found 
an excellent agreement when modelling the SZ emission over their sample, either 
with the universal pressure profile from A10 or with the model from 
\citet{bulbul10}. 
Our SZ and X-ray data are in excellent agreement over the radial range 
(0.1--1)\,$\Rv$, providing tighter observational constraints.

{Our results, together with the SPT, and ACT results, have to be compared with 
other works that reported a difference between SZ and X-ray measurements.
Two papers based on WMAP data reported a measured SZ signal lower than the expected signal from their X-ray properties.  
Based on a {\it WMAP\/} (one year) analysis for 31 randomly selected nearby clusters, \citet{lieu06} found a ratio of one fourth between the measured SZ signal and the expected signal from X-ray constraints. An analysis based on {\it WMAP}-7 data for 49 clusters \citep{kom11} found this ratio to be 0.5--0.7. However, these claims did not agree with other independent {\it WMAP}-5
statistical analysis for 893 clusters \citep{mel11b}.  
Neither do we, obtaining instead an excellent agreement between X-ray and SZ properties within $\Rv$, in particular, when using the \Planck\ frequencies overlapping those of {\it WMAP\/} (see profiles for 30, 44 and 70\,GHz in 
Fig.~\ref{f:freq}). The {\it WMAP\/} fluxes derived by \citet{lieu06} and \citep{kom11} are also discrepant with fluxes from OVRO/BIMA  \citep{bon06}. These differences could be due, for instance, to complex effects involving large-scale beams and/or non-linear gains \citep[e.g.][and references therein]{whitbourn11}.}

\subsubsection{The outskirts}

Beyond $\Rv$, the observational situation is even more wanting. Only a few 
X-ray observations with \xmm, {\it Chandra\/} and/or {\it Suzaku\/}
constrain the density, 
temperature or gas fraction profiles out to ${\sim}\,R_{\rm 200}\simeq1.4\,\Rv$ 
\citep{geo09,rei09,urb11,sim11,walker12,walker12b}. This type of X-ray 
detection remains very challenging and requires very long exposure times, as at 
larger radii the X-ray emission is extremely faint. 
We recall that beyond $\Rv$ the universal pressure profile from A10 was 
constrained by predictions from numerical simulations, not by observations.  
Probing the gas with the SZ effect is therefore a powerful alternative, as 
shown, for instance, by the SPT average emission over 15 clusters \citep{pla10} 
or the {\it WMAP\/}
statistical analysis over ${\sim}\, 700$ clusters by \citet{atrio08}, 
though the latter is affected by the limited resolution and sensitivity of the 
survey. Conversely, from the tentative analysis of {\it WMAP}-3 data for $193$ 
clusters with X-ray temperature above 3\,keV, \citet{afs07} provided 
constraints for the cluster pressure profile (significant out to ${\sim}\, 
1.3\,\Rv$). In agreement with these earlier works, the \Planck\ measurements of 
the present study are the first to allow a precise description of the thermal 
pressure distribution out to the cluster outskirts.

As shown in App.~\ref{s:ind}, \Planck\ also resolves some individual pressure 
profiles. The work presented by the \citet{planck2012-X} on the Coma cluster is 
even more striking, as the SZ profile detection reaches beyond $3\times\Rv$, 
i.e., as far as the  statistical measurement presented in this work. The 
derived best fit model for the \Planck\ observations of Coma is in full 
agreement with our stacked result.

{While in the inner parts the profiles of CC and non-CC clusters are significantly different, and although the CC profiles lie just below the non-CC beyond $\sim 1\, \Rv$, the two  are compatible within our statistical errors. The same is
true when comparing our best-fit models and A10's for CC and non-CC clusters. This suggests that across our sample the average differences  between the outskirts of the two types of clusters are smaller than the scatter between clusters.}

We have also investigated the change in integrated Comptonisation parameter {derived by comparing our best fit of a GNFW model to that of A10. Within a fixed aperture $\Rv$, known from ancillary data, we have computed $\Yv$ for each cluster in our sample and for both sets of parameters (see Eq.~\ref{s:ymes}). The average ratio across the sample between the value of $\Yv$ from our best profile and that of A10's is $1.02 \pm 0.03$. The difference is thus marginal as the two profiles are alike, and consistent with the ratio of 1.01 between the two GNFW parametrisations at $\Rv$.}  This consistency demonstrates the robustness of the previously-published Planck SZ analysis and scaling relations where the A10 profile was adopted as a fiducial model for known $\Rv$ values \citep{planck2011-5.1b, planck2011-5.2a, planck2011-5.2b, planck2012-I,planck2012-III,planck2012-IV}.
The ratio for the values of $Y$ for the two profiles is 1.13 and 1.19 within fixed apertures of $3\,\Rv$ and
$5\,\Rv$ respectively, as expected given the slightly flatter outer slope derived in this work with respect to that of A10.

%
\subsection{Comparison with theoretical predictions }
\label{s:szsim}

%
\subsubsection{Sets of numerical data}
\label{s:numsim}

We have investigated three sets of simulated clusters in order to compare to 
our combined \Planck\ and \xmm\ pressure profile. All are taken from numerical 
simulations of structure formation in a $\Lambda$CDM cosmology:
\begin{enumerate}
\item We first compared to the set of combined simulations used by A10 to 
derive their universal pressure profile together with the \rexcess\ data.
They comprise 93, 14, and 88 simulated clusters with $M>10^{14}\, \msol$, from 
\citet{borgani04}, \citet{nag07} and \citet{piffaretti08}, respectively.
We refer to this set of simulations  as B04+N07+P08 hereafter.
\item The second set is built from the pressure profiles of 64 massive clusters with 
$M>3\times10^{14}\, \msol$ 
from a simulation by \citet{dolag12}. 
\item The third comprises the 40 most  massive clusters (i.e., 
$M>3\times10^{14}\, \msol$) provided from the numerical simulations described
in \citet{battaglia10,battaglia12}. 
\end{enumerate}

All simulations include treatment of radiative cooling, star formation and 
energy feedback from supernova explosions. Simulations by \citet{dolag12} and 
\citet{battaglia10,battaglia12} have prescriptions for AGN feedback.
The different simulation sets use different techniques and different 
implementations of the physical processes. This ensures a fair description of 
up-to-date theoretical predictions, hopefully bracketing the plausible range on 
the thermal pressure profiles distribution in clusters of galaxy. 

We kept the B04+N07+P08 sample as used by A10, even though its mass limit is 
slightly smaller than our ESZ--{\it XMM\/} sample. The higher mass threshold 
for the two other samples enforce the best possible match to our data. As for 
the observed profiles (see Sec.~\ref{s:ppp}), the simulated cluster profiles were renormalised by the 
factor $\langle f(M)\rangle$. The average value for this factor is $1.0$, $1.08$ 
and $1.03$ for the B04+N07+P08, \citet{dolag12} and \citet{battaglia10} 
simulations, respectively.
Finally, we accounted for the differences in the definition of $\Rv$ and $\Mv$ 
between observations and simulations. We corrected from the differences between 
the true mass and the hydrostatic masses which impact the estimation of both 
$\Rv$ and $P_{\rm 500}$ (see Eq.~\ref{e:pm}). We refer to Sec~4.1 of A10 for a 
complete discussion. Assuming a hydrostatic bias 
$r_{\rm cor}=M_{\rm true}/M_{\rm HE}=1.15$ \citep{kay04,piffaretti08},
we divided the 
values of $\Rv$ and $P_{\rm 500}$ derived from the simulations by 
$r_{\rm cor}^{1/3}$ and $r_{\rm cor}^{2/3}$, respectively.

\begin{figure}[!t]
\center
\includegraphics[scale=1.,angle=0,keepaspectratio,width=\columnwidth]{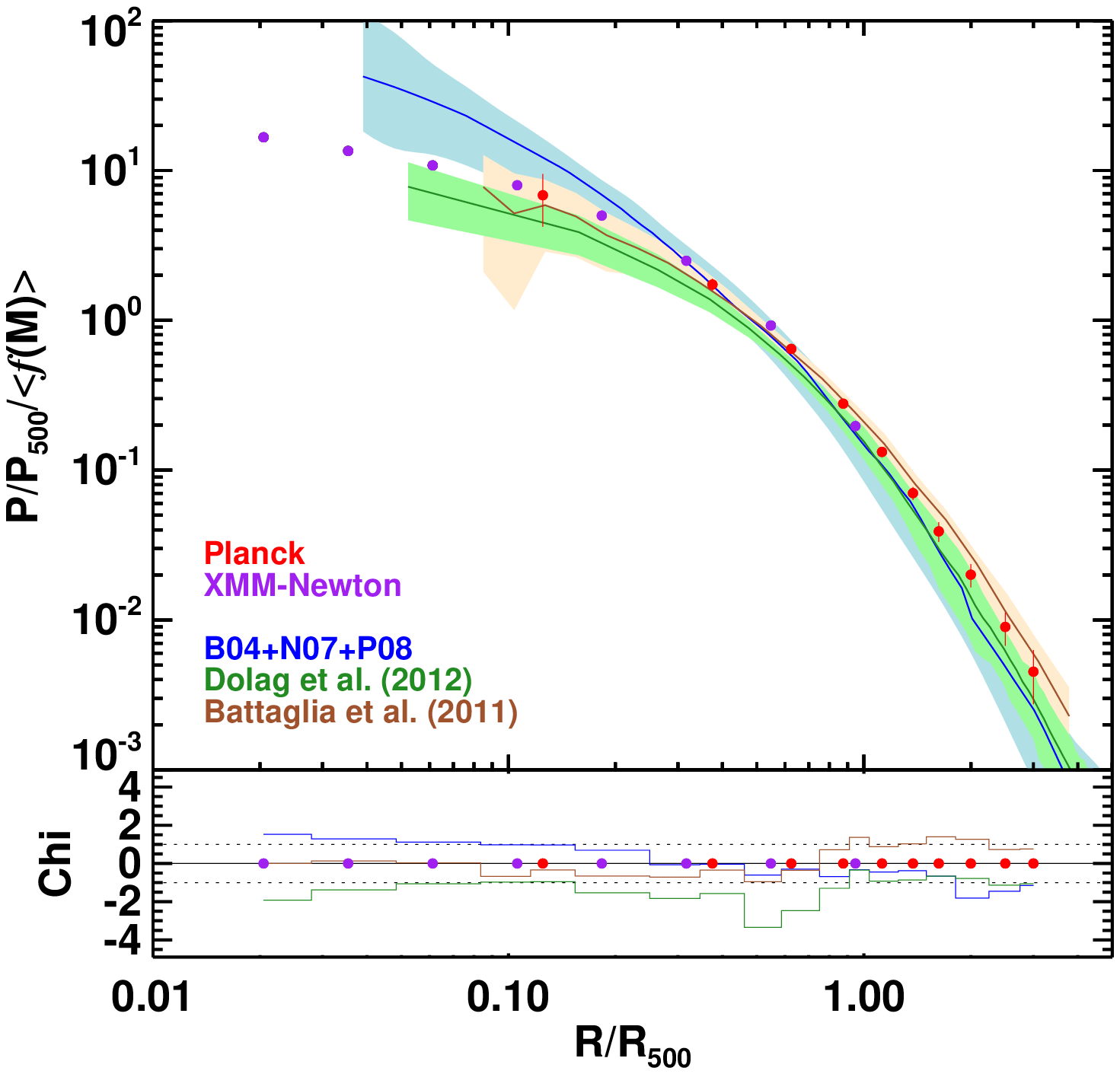}
\caption{\footnotesize{Measured pressure profiles from \Planck\ and \xmm,
displayed as in Fig.~\ref{f:ppp}. The three shaded areas mark the dispersions 
about the average profiles for three samples of simulated clusters: in blue, 
the simulations from \citet{borgani04}, \citet{nag07} and \citet{piffaretti08},
which were used in 
A10 to derive the universal pressure profile together with the \rexcess\ data; 
in green, the simulated sample of clusters from \citet{dolag12}; and in brown 
the simulated clusters
from \citet{battaglia12}. The corresponding average profiles for each set of 
simulations are plotted as solid lines using the same colour scheme. In the 
bottom panel, we present the $\chi$ profiles between the observed profile and 
the simulated average profiles, taking into account their associated dispersion. 
See Sect.~\ref{s:ppp} for more details.}}
\label{f:psim}
\end{figure}

All three sets of data are reported in Fig.~\ref{f:psim}. It is well beyond the 
scope of this paper to extensively discuss the comparison between theoretical 
predictions.  Nevertheless,
we note that they agree within their respective dispersions 
across the whole radial range. The  \citet{dolag12} and  
\citet{battaglia10,battaglia12} profiles best agree within the central part, 
and are flatter than the B04+N07+P08 profile. This is likely due to the 
implementation of AGN feedback, which triggers energy injection at the clusters 
centre, balancing radiative cooling and thus stopping the gas cooling. In the 
outer parts where cooling is negligible, the B04+N07+P08 and \citet{dolag12} 
profiles are in perfect agreement. The \citet{battaglia10} profile is slightly 
higher, but still compatible within its dispersion with the two other sets. 
Here again it might be due to the specific implementation of the simulatons.

\subsubsection{Comparison with the \Planck\ pressure profile}
\label{s:obssim}

Figure~\ref{f:psim} compares our SZ+X-ray observational constraints to the 
theoretical predictions described above. Overall, our observed pressure profile 
lies within the scatter across the profiles from these various sets of  
simulated clusters.

In the central parts, the data points lie on the lower envelope of the scatter 
of the B04+N07+P08 simulations, similarly to the \rexcess\ sample profiles 
(left panel of Fig.~\ref{f:ppp} and figure~7 in A10). This flatter inner slope 
is more compatible with that of the clusters from simulations which include the 
effect of AGN feedback \citep{dolag12,battaglia12}, although our measured 
points lie above them.

With increasing radius (i.e., $R>\Rv$) both SZ and X-ray profiles are 
marginally compatible with the stacked profiles for simulated clusters. 
Interestingly our profile lies above the B04+N07+P08 and \citet{dolag12}
simulated clusters, and below the \citet{battaglia12} clusters.
In the outer parts of halos, the predicted pressure in numerical simulations is 
essentially sustained by gravitational heating. The general agreement (within 
the dispersion) with our measured pressure profile favours the equilibrium
between 
ions and electrons in the outskirts, i.e., both populations have an equivalent 
temperature. Indeed, a difference would result in a drop in thermal 
pressure \citep{rudd09}. This may suggest that we have the correct global
understanding of the intra-cluster thermal pressure support. 
However, the discrepancies between simulations and the average observed
profile in the clusters' outer parts  calls for a deeper 
investigation of the gas distribution and structure in clusters.  Thus we
 need more detailed
modelling of the baryon physics in cluster outskirts, where incomplete 
virialisation of the intra-cluster medium affects the thermal pressure. Even 
if a universal behaviour of cluster quantities is anticipated from the simplest 
model of gravitational collapse, dispersion is also expected, even without 
considering the effects of non-gravitational physics, as shown by numerical 
simulations of halo formation \citep[e.g., ][]{navarro97, reed11, 
bhattacharya11, gao12}. This is plainly reflected in the large level of 
dispersion that we observe across our sample of simulated clusters. 

On the observational side, our investigation of the sub-sample of CC and non-CC 
clusters did not show differences in the outer shape of the pressure profile 
(CC and non-CC cluster profiles are compatible within the measured statistical 
error bars). This might indicate that the aforementioned  dispersion is not 
mainly driven by the dynamical state of clusters, which likely affects 
primarily the inner parts of halos. 
If our \Planck\ constraints over the whole ESZ--{\it XMM\/} sample give a 
flatter outer shape with respect to the average simulated pressure profiles by 
B04+N07+P08 and \citet{dolag12}, we need nonetheless to keep in mind that our 
current sample may be affected by selection biases, as discussed in 
\citet{planck2011-5.2b}.

%
\subsection{Constraint on the gas mass fraction}
\label{s:fgas}
As already mentioned, X-ray measurements hardly reach density contrasts of 
${\sim}\,200$  \citep[e.g.,][]{geo09,rei09,urb11,sim11}, so modelling beyond 
$R_{200}$ only relies on predictions from numerical simulations.  
We used our \Planck\ and \xmm\ derived pressure profile to investigate the 
average gas mass fraction distribution across our sample. Assuming the ICM to 
be a perfect gas, the thermal pressure is the genuine product of the 
temperature and the density, i.e., $P(r)\propto n_{e}(r)\times kT(r)$. 
Following previous X-ray works \citep{pra07,arn10,democles10}, we used the best 
fit to the spectroscopic temperature profile \citep{dem12} with the analytical 
function proposed by \citet{vik06} for each of our clusters. Beyond the reach 
 of the X-ray observations (i.e., [0.5--1.2]$\,\Rv$), we have worked with two 
hypotheses for the temperature: (H1) the extrapolation of the best fit model (note tha ); or 
(H2) a constant temperature fixed to the last radial bin value. We have then 
derived the density profiles, and integrated them over the cluster volume to 
obtain the gas mass radial distribution. Meanwhile, we modelled each total mass 
profile with an NFW profile with the mass, $\Mv$, derived from Eq.~\ref{e:yx}, 
and the concentration estimated from the $c$-$-M$ relation by 
\citet{bhattacharya11}. We checked that the average of the individual NFW mass 
models  agrees with the NFW model derived from the average values of $\Mv$ and 
$c_{\rm 500}$ across the sample.  
The average gas fraction profile is computed as the ratio of the average gas 
mass model and the total mass  model profiles. 

\begin{figure}[!t]
\center
\includegraphics[scale=1.,angle=0,keepaspectratio,width=\columnwidth]{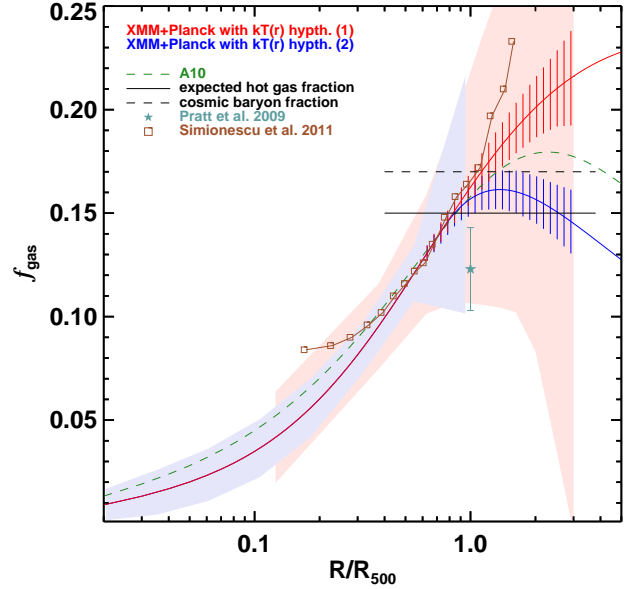}
\caption{\footnotesize{Gas mass fraction profile derived from the combined 
\Planck\ and \xmm\ pressure profile, assuming for the temperature profile: (H1) 
the average best fit model across the sample from X-ray spectroscopy (red line 
and striped area); or (H2) the same but extrapolating beyond $\Rv$ to a constant 
value equal to the average temperature measured in the last radial bin across 
the sample (blue line and striped area). The green dashed curve marks the 
expected gas fraction profile assuming the A10 pressure profile and $kT(r)$ as 
in hypothesis (1). The star gives $f_{\rm gas}(r=\Rv)$ for \rexcess\ 
clusters with $\Mv > 5\times 10^{14}\,\msol$ \citep{pra09}. In maroon we 
reproduce the gas mass fraction profile derived from {\it Suzaku\/}
measurements for the Perseus cluster \citep{sim11}. The solid and dashed black lines mark the 
cosmic baryon fraction expected from CMB measurements \citep{kom11} and the 
expected gas fraction, assuming that 12\% of baryons are in stars, respectively.
The shaded blue and red areas translate the dispersion in the pressure profile 
across the ESZ--{\it XMM\/} sample as shown on Fig.~\ref{f:ppp}, assuming 
hypothesis (H1) for $kT(r)$. See Sec.~\ref{s:fgas}}}
\label{f:fgas}
\end{figure}

Figure~\ref{f:fgas} shows the resulting gas mass fraction distribution for the 
two hypotheses on the temperature at large radii (red and blue solid lines). 
The red and blue striped areas picture the statistical errors measured from the 
SZ+X-ray pressure profile.  The red and blue shaded areas overplot the \xmm\ 
and \Planck\ dispersion onto the pressure profile (see Fig.~\ref{f:ppp}) using
hypothesis (H1) for the temperature, and provide an illustration of the 
cluster-to-cluster variation of $f_{\rm gas}$ within our sample.

From the stacked extrapolated temperature models, the average temperature is 
{$1.3$\,keV at $3\, \Rv$, which leads to $f_{\rm gas}\,{\sim}\,0.2$. An 
underestimation of} this temperature by a third (i.e., $kT\,{\sim}\,1$\,keV)
would increase the gas fraction likewise (i.e., $f_{\rm gas}\,{\sim}\,0.3$).
For the lower bound, hypothesis (H2) fixes the temperature of the averaged
profile beyond $\Rv$ to ${\sim}\, 4$\,keV, which likely overestimates the
true average value of 
the temperature across our sample in the (2--3)$\,\Rv$ radial range. Therefore
we are confident that both hypotheses fairly bracket the range for the gas 
fraction distribution in massive clusters out to $3\,\Rv$. {In other words, our constraints on the pressure profile imply that the temperature required to flatten  the gas fraction profile at $3\, \Rv$ to the expected cosmic value lies between 1 and 4~keV and is likely closer to the lower value. }
With hypothesis (H1) for $kT(r)$, the green dashed line gives the gas fraction  
for the A10 pressure profile. As for our SZ+X-ray pressure profile, it leads to 
$f_{\rm gas}$ values above the expected gas fraction out to $3\,\Rv$.

At $\Rv$, the measurement of $f_{\rm gas}$ for massive clusters presented by 
\citet{pra09} (see also \citealt{giodini09}, for massive clusters with 
(4--10)$\times 10^{14}\,\msol$) is lower by a factor 1.3 with respect to our 
estimate (although compatible within $\pm2\sigma$).
We can also compare our result to the gas fraction profile reported by 
\citet{sim11} using {\it Suzaku\/} measurements
in the direction of the Perseus cluster. {The derived gas fraction profile is in very good agreement with ours within $\Rv$. Beyond this radius, it rises within our plausible dispersion envelope (red and blue shaded areas on 
Fig.~\ref{f:fgas}). Note however, that such a time consuming X-ray observation sampled less than 5\% of the 
sky area covered by the cluster and as a specific measurement is neither representative 
nor out of the allowed range for the gas fraction in cluster outskirts.}

The constrained interval for the gas fraction from our observed pressure 
profile encompasses the cosmic gas fraction derived from CMB measurements 
\citep[solid black line on Fig.~\ref{f:fgas}, from {\it WMAP}-7 results by ]
[]{kom11} 
and the expected hot gas fraction (dashed black line) assuming that 12\% of
cluster baryons are in stars \citep{gonzalez07,giodini09}.

%
\section{Conclusion}
\label{s:con}
From the \Planck\ nominal mission (i.e., 14 months of survey), making 
use of its full spectral coverage of the SZ spectrum, we have extracted and 
reconstructed the SZ signal distribution in clusters of galaxies for a sample 
of 62 massive nearby clusters. All are individually detected in the survey
with high significance \citep{planck2011-5.1a} and were previously used to 
investigate the total integrated SZ flux and the SZ scaling relations 
\citep{planck2011-5.2b}.  We have scaled and averaged together all the SZ 
profiles in the sample into a stacked profile. We have statistically detected 
the SZ signal out to $3\times \Rv$, providing, for the first time, stringent 
observational constraints on the ICM gas beyond $\Rv$ and out to a density 
contrast of $\delta\,{\sim}\,50$--100. 

From the \Planck\ SZ raw profiles, we have derived the underlying thermal 
pressure profiles of the ICM. Together with the pressure 
profile derived from the \xmm\ data, we have provided for the first time a 
comprehensive observational view of the distribution of thermal pressure 
support in clusters from $0.01$ out $3\times\Rv$. We have fitted these unique 
measurements to a generalised NFW profile. Our best analytical representation 
over this wide radial range is given by the parameters $[P_0, c_{\rm 500}, 
\gamma, \alpha, \beta]= [6.41, 1.81, 0.31, 1.33, 4.13]$.

Our observational measurements further confirm the agreement of the SZ and 
X-ray constraints on the intra-cluster gas properties within the inner part of 
the clusters commonly considered to be virialised, i.e., $\Rv$, as found in
\citet{planck2011-5.2a} and \citet{planck2011-5.2b}.
Overall it also agrees with a wide range of  simulated clusters through the whole radial range, although in the 
central regions it matches best the numerical predictions that implement {prescriptions} for AGN 
feedback. The statistical nature of our stacked detection at large radii 
provides the average trend for the thermal pressure distribution in cluster 
outskirts out to $3\times \Rv$, which is slightly flatter than most theoretical 
predictions.

In conjunction with X-ray constraints on the temperature profile of our  clusters, 
we have derived the profile of the gas mass fraction out to the 
cluster outskirts. From reasonable hypotheses on the gas temperature to account 
for the lack of constraints beyond $\Rv$, we have bracketed a range for the gas 
fraction in the cluster outer regions, which is compatible with the cosmic 
baryon fraction and the expected gas fraction in halos.%

The processes governing the thermodynamical state of the outer regions in 
clusters still need to be understood from the theoretical and  observational 
points of view \citep[see ][for a review and references]{kra12}. 
Issues such as gas clumping will affect the pressure estimation \citep[i.e., 
effect on the X-ray surface brightness, see ][]{ron06}, departures from
hydrostatic equilibrium and contribution from non-thermal pressure (due to 
magnetic fields and/or cosmic rays) will modify the gas  fraction. 
In this regard  SZ observations provide a straightforward description of the 
thermal pressure distribution in massive halos, and  a clear path to the gas 
fraction determination.
{Future SZ instruments with increased spatial resolution and sensitivity, and retaining the ability to map clusters of galaxies out to large radii will certainly provide us with further  details and insight. In the meantime,}
  with \Planck\ we bring unique observational constraints that are 
extremely valuable to further test and understand the physics at play in the
outskirts of clusters. 
%
\begin{acknowledgements}
The development of \Planck\ has been supported by: ESA; CNES and 
CNRS/INSU-IN2P3-INP (France); ASI, CNR, and INAF (Italy); NASA and DoE (USA); 
STFC and UKSA (UK); CSIC, MICINN and JA (Spain); Tekes, AoF and CSC (Finland); 
DLR and MPG (Germany); CSA (Canada); DTU Space (Denmark); SER/SSO 
(Switzerland); RCN (Norway); SFI (Ireland); FCT/MCTES (Portugal); and DEISA 
(EU). We acknowledge the use of the Healpix software.

\end{acknowledgements}

\bibliographystyle{aa}
\bibliography{Planck2011-ppp,Planck_bib}

\appendix 
\section{SZ map reconstruction methods}
\label{s:met}

In this Appendix we discuss the different SZ map reconstruction methods applied 
to the \Planck\ data in Sect.~\ref{s:ilc}.

\begin{enumerate}
\item {\bf MILCA} \citep{hurier10}: The thermal SZ signal reconstruction is 
performed on the six \Planck\ all-sky maps from 100\,GHz to 857\,GHz. MILCA 
(Modified Internal Linear Combination Algorithm) is a component separation 
approach aiming at extracting a chosen component (here the thermal SZ signal) 
from a multi-channel set of input maps. It is mainly based on the well known 
Internal Linear Combination approach \citep[e.g.,][]{eriksen04}, that searches 
for the linear combination of the input maps that minimises the variance of the 
final reconstructed map by imposing spectral constraints. In this paper, we 
applied MILCA using two constraints: preservation of the thermal SZ signal 
(knowing the SZ spectral signature); and removal of the CMB contamination in the 
final SZ map (also making use of the well known spectrum of the CMB). In 
addition, to compute the weights of the linear combination, we have used the 
extra degrees of freedom to minimise residuals from other components (2 
degrees) and from the noise (2 degrees). The noise covariance matrix was 
estimated from the frequency error maps (see Sect.~\ref{s:data}).

\item {\bf NILC}: Needlet ILC performs a linear combination of the observed 
maps which has minimum variance under the constraint of offering unit response 
to the component of interest (here the thermal SZ, whose frequency scaling is 
known). The weights of the ILC depend on the covariance between the various 
observations, and can be computed, for example, on domains of the observed 
pixels (pixel space) or angular scales (spherical harmonics).
In the case of NILC, 
covariances (and hence weights for component separation) are computed 
independently in domains of a needlet decomposition (spherical wavelet frame). 
The needlet decomposition provides localisation of the ILC filters both in 
pixel and in multipole space, allowing us to deal with local contamination 
conditions varying both in position and in scale. NILC was developed to extract
a CMB map from {\it WMAP\/} data \citep{delabrouille09} and was also tested for
SZ effect 
extraction in \citet{leach08}. Multi-component extensions have been investigated 
by \citet{remazeilles11}.

\item {\bf GMCA} \citep{bob08} is a blind source separation method developed
for separating sources from instantaneous linear mixtures. The components are 
assumed to be sparsely represented (i.e., have a few significant samples in a 
specific basis) in a so-called sparse representation $\Psi$ (typically
wavelets). The assumption that the components have a sparse representation in
the wavelet domain is equivalent to assuming that most components have
a certain 
spatial regularity. These components and their spectral signatures are then 
recovered by minimising the number of significant coefficients in  $\Psi$. 
Recently, L-GMCA has been further introduced to analyse the CMB data in a local 
and multi-scale manner \citep{bob12}. More precisely, the multi-channel data
are analysed in four frequency bands in spherical harmonics (i.e., wavelet
bands).  In each wavelet band, GMCA is applied locally on small patches with 
band-dependent sizes. In each band, the observations are analysed at the same 
band-dependent resolution. In \citet{bob12}, it was shown this local analysis 
approach enhances the separation quality. The spectral signatures of CMB, 
free-free and SZ are assumed to be known. 

\end{enumerate}

\begin{figure}[!thhh]
\center
\includegraphics[scale=1.,angle=0,keepaspectratio,width=\columnwidth]{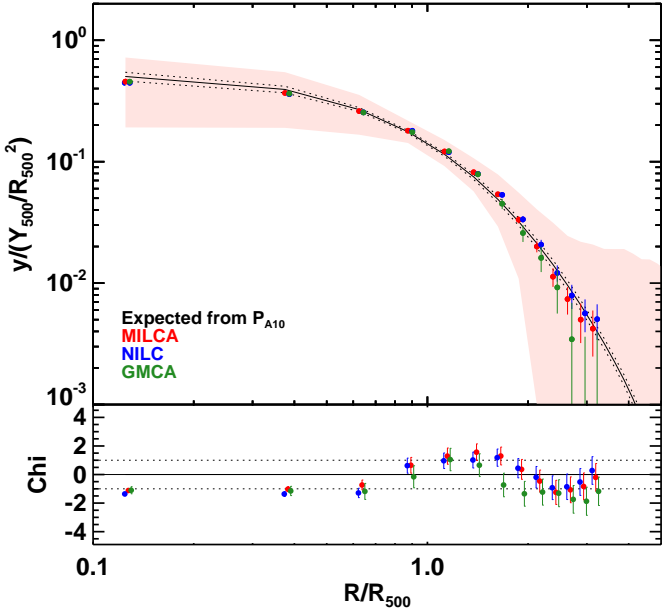}
\caption{\footnotesize{Comparison between the SZ profile reconstruction from 
MILCA, NILC and GMCA 
The comparison is presented in the form of the averaged stacked radial profile 
from our 62 clusters of the ESZ--{\it XMM\/} sample, individually scaled 
respectively in the $x$- and $y$-axis directions according to $\Rv$ and $\Phi$ 
(Eq~\ref{e:phi}). The light-red shaded area marks the dispersion about the 
average stacked profile for the MILCA method.  The points within each
individual profile are correlated at about the ${\sim}\,20$\% level
(see Sect.~\ref{s:psz}) and the plotted errors correspond to the square root of 
the diagonal elements of the covariance matrix of the profile. 
The solid black line (labelled in the legend as 'Expected from $P_{A10}$') is 
the average stacked profile obtained from the expected SZ individual profiles 
drawn from the universal pressure profile by A10, parameterised according $\Rv$ 
and $\Yv$ derived from the \xmm\ data analysis presented in 
\citet{planck2011-5.2b}). The two dotted lines indicate the associated 
dispersion across the sample.
The bottom panel shows the value of $\chi$ at each point of the observed 
profiles with respect to the expectations from the universal profile of A10.}}
\label{f:metapx}
\end{figure}

%
\section{Validation on simulations}
\label{s:sim}

\subsection{Profiles}

We have validated our reconstruction of the SZ signal and profile computation 
methods with simulations. We used the \Planck\ sky model \citep[PSM, 
][]{delabrouille12} to simulate the whole sky as seen by \Planck. The PSM 
includes most astrophysical components acting as foreground or background 
contamination with respect to the SZ signal (i.e., Galactic dust, free-free and 
synchrotron emission, extragalactic sources in the infrared and radio parts of 
the spectrum and CMB). For the instrumental noise we added to the simulation 
the all-sky error map drawn from the \Planck\ jack-knife maps at each frequency 
(see Sect.~\ref{s:data}), thus mimicking the noise properties of the actual 
survey (noise and systematic effects). For the SZ component, we built a full 
sky template for all the 1743 clusters from the MCXC meta-catalogue 
\citep{pif11}, which includes all 62 clusters in our sample. Each cluster was 
modelled assuming spherical symmetry and a thermal pressure distribution 
following the universal profile from A10.
We recall that the universal pressure profile provided by these authors is the 
best fit of a generalised NFW profile \citep{nag07} to the  median profile 
derived from the \rexcess\ pressure profile and the predictions from three 
different sets of numerical simulations \citep[i.e., 
from][]{borgani04,nag07,nag07sim,piffaretti08}. 
In our case, the universal pressure profile is  parametrised using  the values 
of $\Rv$ and  $M_{\rm 500}$ from the MCXC, together with the  $M$--$Y_{\rm X}$
scaling relation from \citet{arn07, arn10} to estimate $Y_{\rm 500}$. 

We processed the simulated \Planck\ sky maps through each of our SZ map 
reconstruction methods. For our sample of 62 clusters, we extracted the patches 
and computed the individual profiles as described in Sect.~\ref{s:psz}. We 
stacked the simulated profiles for the three methods after renormalisation,
making use of the aforementioned $\Rv$ and $\Yv$ values from the MCXC 
prescription.

The resulting stacked profiles are displayed in Fig.~\ref{f:metapx}. All three 
tested methods agree remarkably well over the whole radial range. Taking into 
account the correlated errors of each profile, the $\chi^2$ {over the radial 
range (0--10)$\times\Rv$ for NILC and GMCA with respect to MILCA are 
0.78 and 1.54, respectively}.
The reconstructed profiles are also in good agreement with the reconstructed 
stacked model. Accounting for the errors on the model and the correlated errors 
between points on the reconstructed profiles, the reduced $\chi^2$ for MILCA, 
NILC and GMCA with respect to the stacked model profile are 0.80, 0.66 and 1.03 
within $3\times\Rv$. 
This excellent agreement also translates into a small relative error  below 
15\% in the same range of radii.  Beyond $R\,{\sim}\,2\,\Rv$,
the three reconstructed profiles {drop slightly below 
the input model, even though they still agree within their dispersions and 
errors. This drop is caused by an intrinsic over-prescription of point sources 
in the PSM simulations with respect to real data. 
Despite our careful masking, the unidentified or unresolved extra sources in 
the PSM affect the power spectrum of the noise, increasing the background level 
and making it more difficult to detect the SZ signal further out in the 
simulations than in real data. We stress that this is an effect limited to our 
test simulations, and not relevant for the real data}.

\subsection{Consistency between $\Yv$ measurements}
\label{s:ymes}

\begin{figure}[!t]
\center
\includegraphics[scale=1.,angle=0,keepaspectratio,width=\columnwidth]{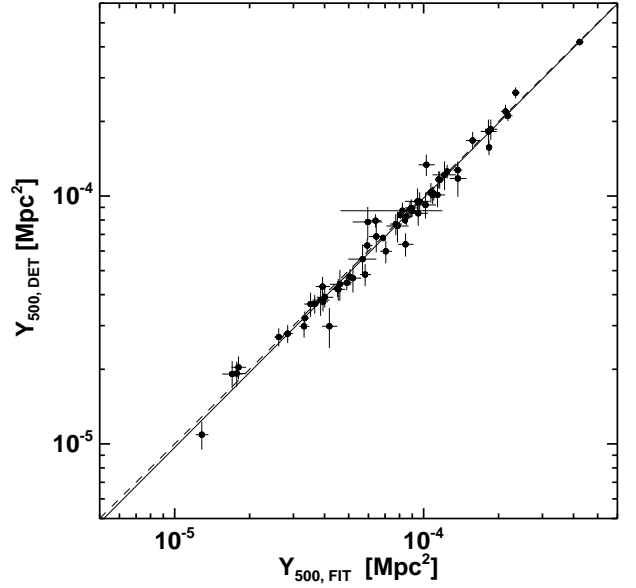}
\caption{\footnotesize{
Comparison of $\Yv$ values from the \Planck\ cluster detection method (MMF3) 
against our best fit for each of our 62 clusters (from NILC in this case). In 
both cases the distribution of thermal pressure is assumed to follow a 
universal pressure profile (A10) for which the scaled radius $\Rv$ is fixed to 
the \xmm\ best fit value. See Sect.~\ref{s:fit} for details.} }
\label{f:fit}
\end{figure}

To further validate the above, and as a consistency check with previous 
\Planck\ results, we fitted each individual SZ profile with a projected, 
PSF-convolved universal pressure profile. We fixed $\Rv$ to the best fitting 
X-ray value from \citet{planck2011-5.2b} and only fitted the normalisation, 
$Y_{\rm 500}$. For a given profile, $P$, with an associated covariance matrix, 
$C$, and a model $M$, the $\chi^2$ statistic can be expressed as
\begin{equation}
\chi^2 = (P-Y_{\rm 500}\times M)^{\rm T}C^{-1}(P-Y_{\rm 500}\times M).
\end{equation}
The solution that minimises $\chi^2$, ${Y}_{\rm 500}$, and its associated 
uncertainty, ${\sigma}_{Y_{500}}$, are analytic:
\begin{equation}
{Y}_{\rm 500} = {\sigma}_{Y_{500}}^2\, M^{\rm T}C^{-1}P; \\
\label{e:ymes}
\end{equation}
\begin{equation}
{\sigma}_{Y_{500}}^2=(M^{\rm T}C^{-1}M)^{-1}.
\end{equation}
We compared the resulting value of $Y_{\rm 500}$ to those obtained from the 
detection algorithms PwS and MMF3. Both detection algorithms were run under the 
same assumptions, i.e., universal pressure profile with the ``non-standard'' 
parameterisation and a fixed size ($\Rv$). The fluxes were derived from the 
\Planck\ nominal mission survey maps. These fluxes are in agreement with those 
extracted from the \Planck\ first year survey \citep{planck2011-5.2b}. For 
instance, for MMF3, the average ratio is $1.04\pm 0.09$, compatible within the 
average relative errors with the fluxes extracted from the nominal 
mission.

The relation between the integrated Comptonisation parameters derived from the 
SZ profiles and from the detection methods is displayed for MMF3 vs NILC fuxes 
in Fig.~\ref{f:fit}. A BCES linear fit accounting for the errors in both $x$-
and $y$-axis
directions \citep{akr96} yields a best fitting slope and normalisation of 
$1.00\pm 0.02$ and $0.01\pm 0.01$ ($\bar{\chi}^2=0.83$), respectively. There is 
4\%  dispersion about this relation.
Furthermore, the median and associated deviation for the one-to-one ratios of 
PwS and MMF3 fluxes compared to the fitted NILC profile values are $1.03\pm 
0.11$ and $0.98\pm 0.08$, respectively. 
The median ratios for the errors are  $1.10\pm 0.36$ are $1.64\pm 0.34$. 
The errors associated with our best fit values are in very good agreement with 
those from PwS; however, they are smaller than those derived from the MMF3 
method. This difference with MMF3 might be explained by the different treatment 
of the noise. {Whilst we estimated the noise from the surroundings of the 
clusters, defined according to the scaled radius of each target, and PwS also 
used a local computation of the noise, the MMF3 algorithm used a larger area to 
characterise the noise properties.}
Our tests against the PwS and MMF3 fluxes emphasise our grasp on the 
characterisation of the local noise for each target, and further assess the 
reliability of our estimates of the statistical errors in our stacked profiles.

Such agreement between the fluxes derived from our fit over the cluster 
profiles and those from the detection algorithm, strongly confirms the 
reliability of our SZ flux estimate and further strengthen the overall result 
stated in the \Planck\ collaboration early papers that the  SZ and the X-ray 
fluxes  within $\Rv$ are fully consistent 
\citep{planck2011-5.1a,planck2011-5.1b,planck2011-5.2a,planck2011-5.2b}.

%
\section{Pressure profiles of individual clusters}
\label{s:ind}
In this appendix we present the individual pressure profiles obtained from 
\Planck\ and \xmm\ data for each of the 62 clusters in our sample (see 
Fig.~\ref{f:ind}). We show the agreement with the universal pressure profile 
(A10) and provide the best fitting model for each cluster to the GNFW profile 
in Table~\ref{t:ind}.  The following hard limit priors were adopted during the fitting procedure: $0<P_0< 100$~; $0<c_{\rm 500}<10$~; $0< \alpha<10$~; and $0 < \beta < 15$.
\newpage

\begin{figure*}[p]
\center
\subfloat[][]{\includegraphics[angle=0,keepaspectratio,width=1.8\columnwidth]
 {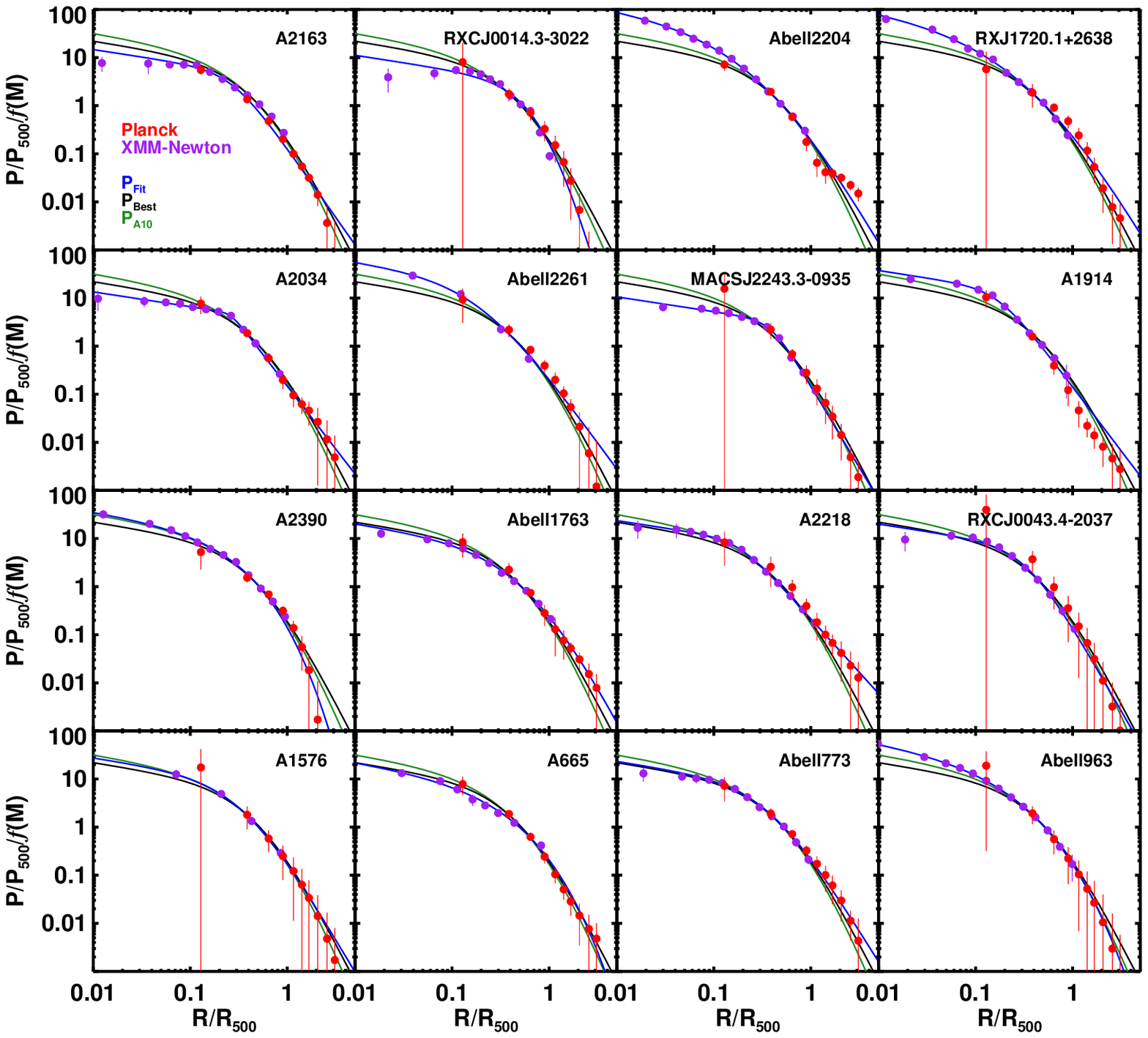}}
\caption{\footnotesize{Measured individual pressure profiles for the
ESZ--{\it XMM\/} sample using the \Planck\ and \xmm\ data. As for the stacked 
pressure profile, the maximum difference point-to-point between the three SZ 
reconstruction methods have been taken into account in the \Planck\ 
measurement error budget  (Sect.~\ref{s:fit}). The best fit model on each 
individual profile is shown as a blue solid line. The black and green solid 
curves mark the  best fit model to the stacked SZ+X-ray pressure profile 
(Sect.~\ref{s:ppp}) and the A10 best fit model, respectively}}
\label{f:ind}
\end{figure*}

\begin{figure*}[p]
\ContinuedFloat 
\center
\subfloat[][]{\includegraphics[angle=0,keepaspectratio,width=1.8\columnwidth]
 {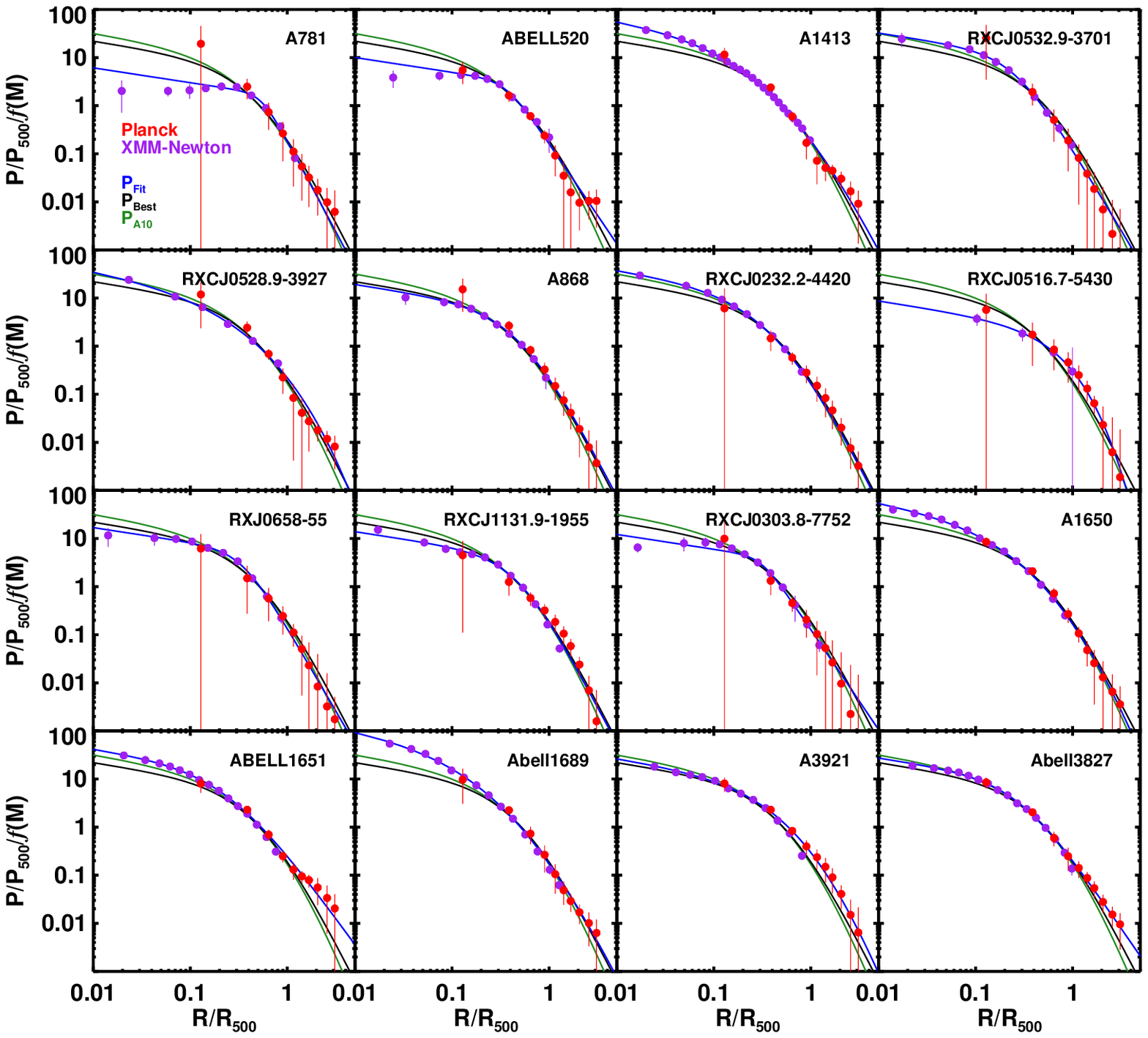}}
\caption{continued.}
\label{f:ind}
\end{figure*}

\begin{figure*}[p]
\ContinuedFloat 
\center
\subfloat[][]{\includegraphics[angle=0,keepaspectratio,width=1.8\columnwidth]
 {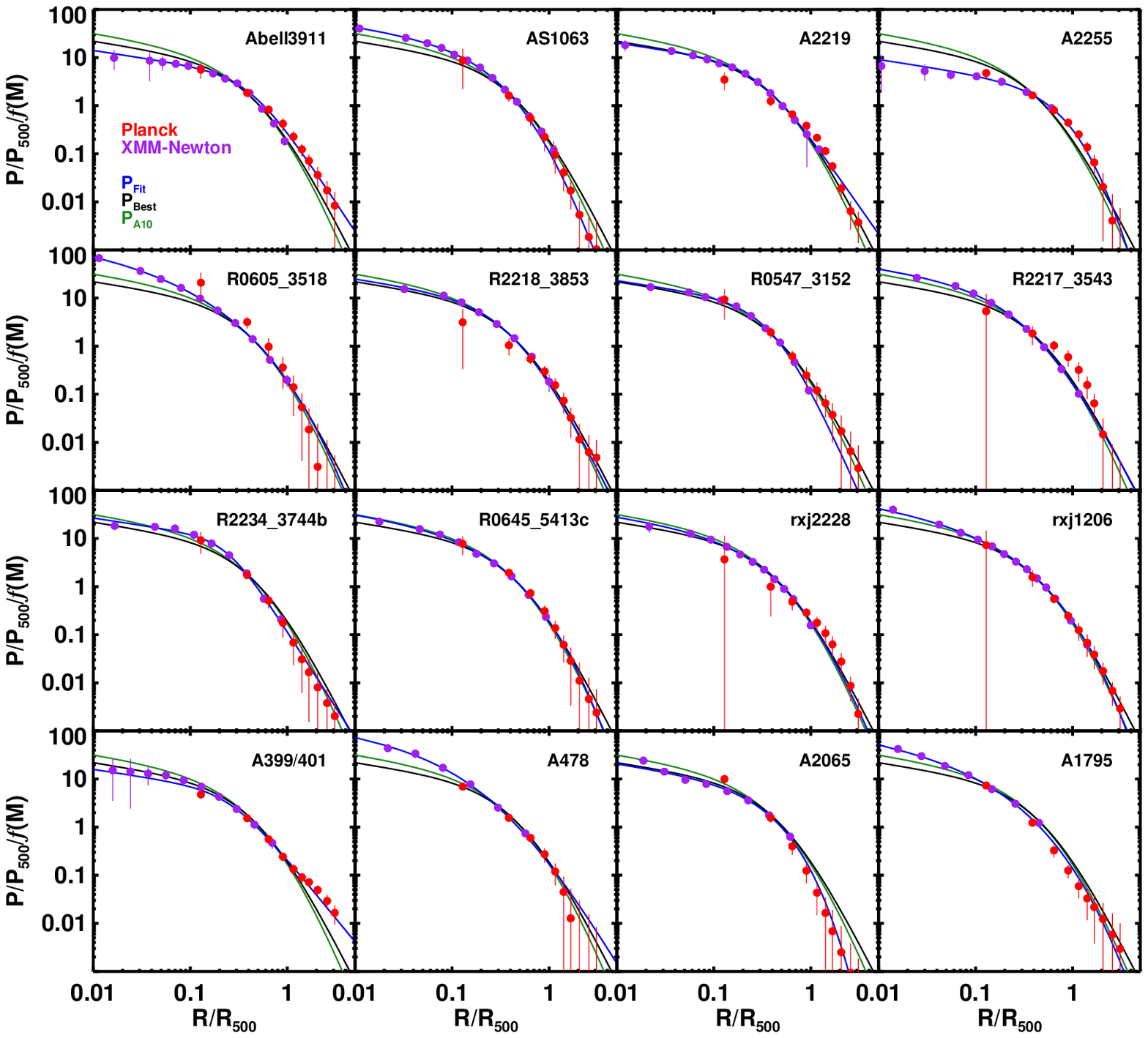}}
\caption{continued.}
\label{f:ind}
\end{figure*}

\begin{figure*}[p]
\ContinuedFloat 
\center
\subfloat[][]{\includegraphics[angle=0,keepaspectratio,width=1.8\columnwidth]
 {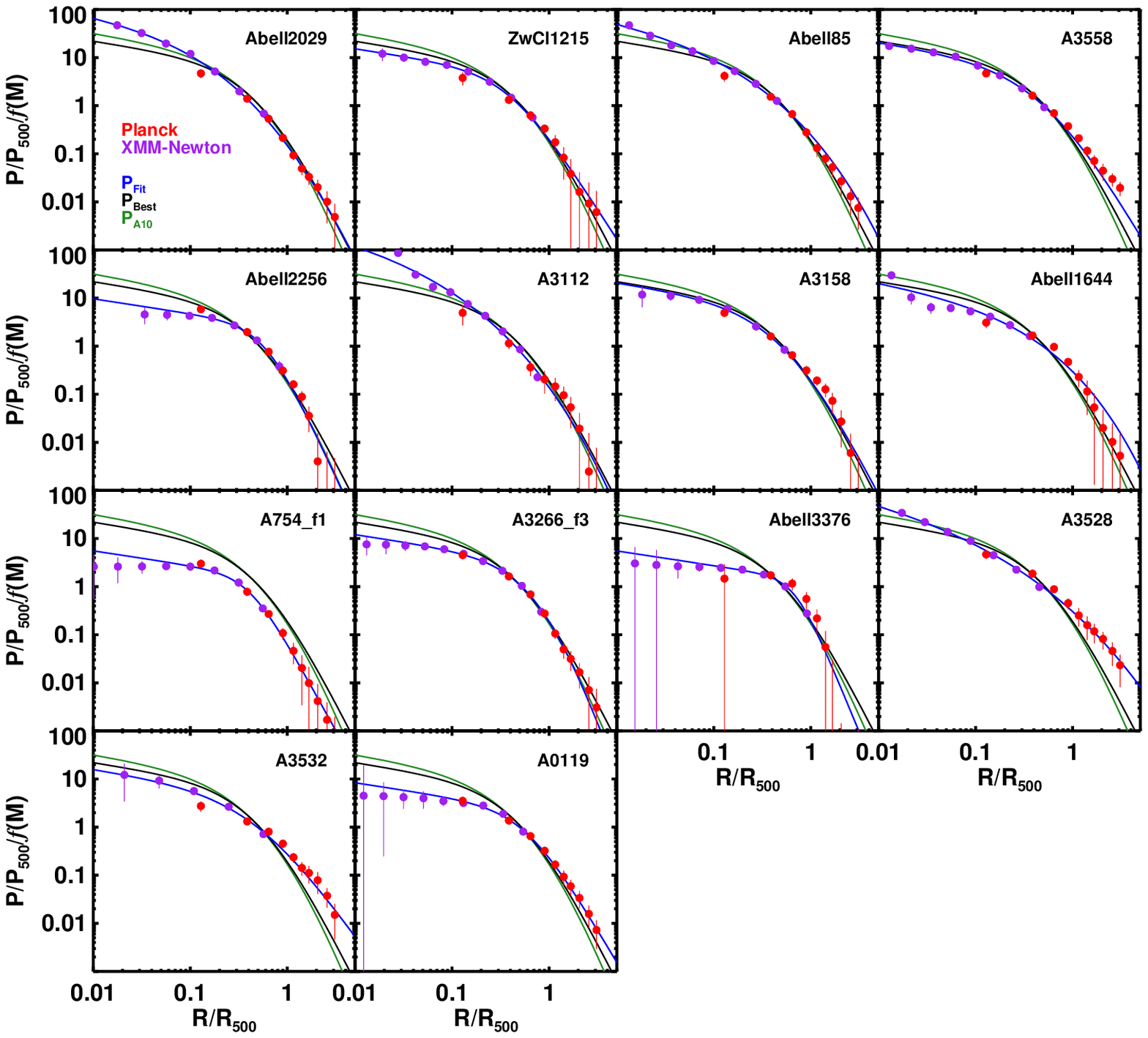}}
\caption{continued.}
\label{f:ind}
\end{figure*}

\begin{table*}[p]
\begingroup
\newdimen\tblskip \tblskip=5pt
\caption{Best fit parameters for a generalised NFW pressure profile for 
individual clusters.}
\label{t:ind}
\nointerlineskip
\footnotesize
\setbox\tablebox=\vbox{
   \newdimen\digitwidth 
   \setbox0=\hbox{\rm 0} 
   \digitwidth=\wd0 
   \catcode`*=\active 
   \def*{\kern\digitwidth}
   \newdimen\signwidth 
   \setbox0=\hbox{+} 
   \signwidth=\wd0 
   \catcode`!=\active 
   \def!{\kern\signwidth}
\halign{\hbox to 1.5in{#\leaderfil}\tabskip 2.4em&
\hfil#\hfil&
\hfil#\hfil&
\hfil#\hfil&
\hfil#\hfil&
\hfil#\hfil\tabskip 0pt\cr
\noalign{\doubleline\vskip 0pt}
\omit\hfil Cluster name\hfil& $P_0$& $c_{500}$& $\gamma$& $\alpha$& $\beta$\cr
\noalign{\vskip 3pt\hrule\vskip 4pt}
A2163&              $*5.28$& $3.64$& $0.31$& $*2.51$& $*2.87$\cr
RXCJ0014.3$-$3022&  $*2.74$& $1.05$& $0.31$& $*1.83$& $*7.24$\cr
A2204&              $35.29$& $2.79$& $0.31$& $*0.83$& $*3.68$\cr
RXJ1720.1+2638&     $30.51$& $2.15$& $0.31$& $*0.74$& $*3.72$\cr
A2034&              $*4.92$& $3.82$& $0.31$& $10.00$& $*2.62$\cr
A2261&              $22.88$& $4.97$& $0.31$& $*1.25$& $*2.79$\cr
MACSJ2243.3$-$0935& $*3.41$& $2.53$& $0.31$& $*4.27$& $*3.38$\cr
A1914&              $15.57$& $5.83$& $0.31$& $*2.75$& $*2.66$\cr
A2390&              $*5.96$& $0.26$& $0.31$& $*0.90$& $14.94$\cr
A1763&              $*5.61$& $1.43$& $0.31$& $*1.10$& $*4.00$\cr
A2218&              $*9.65$& $5.51$& $0.31$& $*2.99$& $*2.23$\cr
RXCJ0043.4$-$2037&  $*6.72$& $2.92$& $0.31$& $*2.15$& $*3.52$\cr
A1576&              $*9.23$& $2.81$& $0.31$& $*1.45$& $*3.45$\cr
A665&               $*3.13$& $0.14$& $0.31$& $*0.80$& $14.38$\cr
A773&               $*7.62$& $2.60$& $0.31$& $*1.46$& $*3.29$\cr
A963&               $*8.82$& $0.17$& $0.31$& $*0.71$& $13.04$\cr
A781&               $*1.78$& $1.82$& $0.31$& $*5.56$& $*3.77$\cr
A520&               $*3.31$& $2.75$& $0.31$& $*4.00$& $*2.98$\cr
A1413&              $17.94$& $1.62$& $0.31$& $*0.83$& $*4.31$\cr
RXCJ0532.9$-$3701&  $11.83$& $3.81$& $0.31$& $*2.00$& $*3.39$\cr
RXCJ0528.9$-$3927&  $*4.62$& $0.07$& $0.31$& $*0.64$& $14.74$\cr
A868&               $*5.65$& $1.88$& $0.31$& $*1.48$& $*3.94$\cr
RXCJ0232.2$-$4420&  $11.50$& $1.95$& $0.31$& $*1.10$& $*4.26$\cr
RXCJ0516.7$-$5430&  $*1.36$& $0.24$& $0.31$& $*1.21$& $14.75$\cr
RXJ0658$-$55&       $*5.69$& $2.92$& $0.31$& $*2.91$& $*3.46$\cr
RXCJ1131.9$-$1955&  $*4.18$& $2.03$& $0.31$& $*1.90$& $*3.85$\cr
RXCJ0303.8$-$7752&  $*4.23$& $3.24$& $0.31$& $*4.94$& $*2.97$\cr
A1650&              $14.21$& $0.78$& $0.31$& $*0.78$& $*6.17$\cr
A1651&              $15.87$& $3.89$& $0.31$& $*1.23$& $*2.81$\cr
A1689&              $33.95$& $1.76$& $0.31$& $*0.77$& $*4.49$\cr
A3921&              $*6.09$& $0.75$& $0.31$& $*0.96$& $*5.63$\cr
A3827&              $*9.74$& $3.42$& $0.31$& $*1.54$& $*2.97$\cr
A3911&              $*4.39$& $2.30$& $0.31$& $*1.98$& $*3.07$\cr
AS1063&             $11.80$& $1.30$& $0.31$& $*1.08$& $*6.18$\cr
A2219&              $*7.04$& $3.25$& $0.31$& $*1.89$& $*2.90$\cr
A2255&              $*1.82$& $0.53$& $0.31$& $*1.41$& $*8.35$\cr
R0605\_3518&        $11.25$& $0.07$& $0.31$& $*0.58$& $14.91$\cr
R2218\_3853&        $*7.51$& $1.97$& $0.31$& $*1.35$& $*4.21$\cr
R0547\_3152&        $*7.13$& $2.22$& $0.31$& $*1.77$& $*4.59$\cr
R2217\_3543&        $13.20$& $2.30$& $0.31$& $*1.12$& $*3.97$\cr
R2234\_3744b&       $*9.98$& $4.10$& $0.31$& $*2.66$& $*3.16$\cr
R0645\_5413c&       $*6.54$& $0.49$& $0.31$& $*0.88$& $*8.03$\cr
RXJ2228&            $*6.92$& $0.92$& $0.31$& $*0.99$& $*5.85$\cr
RXJ1206&            $*6.29$& $0.12$& $0.31$& $*0.70$& $14.80$\cr
A401&               $*5.80$& $3.79$& $0.31$& $*2.08$& $*2.46$\cr
A478&               $30.40$& $3.00$& $0.31$& $*0.84$& $*3.53$\cr
A2065&              $*3.73$& $0.35$& $0.31$& $*1.09$& $15.00$\cr
A1795&              $*8.11$& $0.10$& $0.31$& $*0.63$& $14.98$\cr
A2029&              $21.48$& $0.91$& $0.31$& $*0.66$& $*5.29$\cr
ZwCl1215&           $*4.88$& $2.46$& $0.31$& $*1.65$& $*3.17$\cr
A85&                $*5.99$& $0.02$& $0.31$& $*0.48$& $14.97$\cr
A3558&              $*6.04$& $1.77$& $0.31$& $*1.12$& $*3.58$\cr
A2256&              $*2.72$& $1.65$& $0.31$& $*2.41$& $*4.38$\cr
A3112&              $24.16$& $0.03$& $0.31$& $*0.44$& $14.11$\cr
A3158&              $*5.93$& $1.63$& $0.31$& $*1.17$& $*4.11$\cr
A1644&              $*2.08$& $0.03$& $0.31$& $*0.60$& $14.89$\cr
A754&               $*1.76$& $2.42$& $0.31$& $*2.63$& $*3.66$\cr
A3266&              $*3.05$& $1.15$& $0.31$& $*1.55$& $*5.60$\cr
A3376&              $*1.49$& $1.42$& $0.31$& $*3.57$& $*4.89$\cr
A3528s&             $*5.72$& $0.01$& $0.31$& $*0.36$& $11.63$\cr
A3532&              $*4.79$& $1.90$& $0.31$& $*1.08$& $*2.94$\cr
A0119&              $*2.38$& $1.67$& $0.31$& $*1.81$& $*3.44$\cr
\noalign{\vskip 5pt\hrule\vskip 4pt}}}
\endPlancktablewide
\endgroup
\end{table*}

\raggedright
\end{document}